\documentclass[sigconf, nonacm]{acmart}

\newcommand\vldbdoi{XX.XX/XXX.XX}
\newcommand\vldbpages{XXX-XXX}
\newcommand\vldbvolume{14}
\newcommand\vldbissue{1}
\newcommand\vldbyear{2020}
\newcommand\vldbauthors{\authors}
\newcommand\vldbtitle{\shorttitle} 
\newcommand\vldbavailabilityurl{URL_TO_YOUR_ARTIFACTS}
\newcommand\vldbpagestyle{plain} 

\usepackage[utf8]{inputenc}
\usepackage[english]{babel}
\usepackage[T1]{fontenc}
\usepackage{comment}

\usepackage{graphicx} % Required for inserting images
\usepackage{amsmath,amsfonts}
\usepackage{amsthm}
\usepackage{xcolor}
\usepackage[ruled,vlined,boxed,linesnumbered]{algorithm2e}
\usepackage{caption}
\usepackage{subcaption}
\usepackage{pifont}
\usepackage{multirow}
\usepackage{booktabs}
\usepackage{todonotes}
\usepackage{cleveref}
\usepackage{mathtools}
\usepackage{multirow}
\usepackage{makecell}

\title{Approximate \texorpdfstring{$2$}{2}-hop neighborhoods on incremental graphs: \\ An efficient lazy approach}

\date{\today}

\newcommand{\Prob}[2]{\mathbb{P}_{#1} \left( #2 \right)}
\newcommand{\Expec}[2]{\mathbb{E}_{#1} \left[ #2 \right]}

\newcommand{\bigO}{\mathbf{O}}

\newcommand{\bd}{\Delta} % black degree
\newcommand{\rd}{\delta} % red degree

\newcommand{\lbdd}{b}
\newcommand{\lbddt}{\beta}
\newcommand{\lrd}{\ell}
\newcommand{\lrdr}{\hat{\ell}}

\newcommand{\neigh}{\mathcal{N}} % neighbourhood
\newcommand{\ball}{B} % ball
\newcommand{\lset}{L} % \lset_h(u) is distance h level set wrt u: set of vertices at distance exactly h from u
\newcommand{\apxball}{\hat{B}} % approximate ball
\newcommand{\ok}[1]{locally $#1$-sparse}% ok grpahs with paramater
\newcommand{\lazyscheme}{\textsc{Lazy-Alg}}
\newcommand{\gammaok}{locally $\gamma$-sparse} % gamma-ok
\newcommand{\jacc}{\mathbf J } % Jaccard similarity
\newcommand{\cost}{T}
\newcommand{\rank}{\mathrm{\pi}}

\newcommand{\ie}{i.e., }
\newcommand{\rem}[1]{\todo[inline,color=yellow]{#1}}

\def\dynG{{\mathcal G}}

\newtheorem{theorem}{Theorem}
\newtheorem{corollary}[theorem]{Corollary}

\newtheorem{definition}{Definition}[section]

\newtheorem{fact}{Fact}
\newtheorem{remark}{Remark}

\newtheorem{lemma}{Lemma}

\begin{document}

%%
%% The "author" command and its associated commands are used to define the authors and their affiliations.
\author{Luca Becchetti}
\affiliation{%
    \institution{\textit{Sapienza} University of Rome}
    \city{Rome}
    \country{Italy}
}
\email{becchetti@diag.uniroma1.it}

\author{Andrea Clementi}
\affiliation{%
    \institution{ \textit{Tor Vergata} University of Rome}
    \city{Rome}
    \country{Italy}
}
\email{clementi@mat.uniroma2.it}
%\orcid{0000-0002-1825-0097}
 
% \institution{The Th{\o}rv{\"a}ld Group}
   
\author{Luciano Gualà}
%\orcid{0000-0001-5109-3700}
\affiliation{%
  \institution{ \textit{Tor Vergata} University of Rome}
  \city{Rome}
   \country{Italy}
   }
   \email{guala@mat.uniroma2.it}

\author{Luca Pepè Sciarria}
\authornotemark[0]
\authornote{This work has been supported by the Spoke 1 ``FutureHPC \& BigData'' of the Italian Research Center on High Performance Computing, Big Data and Quantum Computing (ICSC) funded by MUR Missione 4 Componente 2 Investimento 1.4: Potenziamento strutture di ricerca e creazione di ``campioni nazionali'' di R\&S (M4C2-19) - Next Generation EU (NGEU).}
\affiliation{%
  \institution{ \textit{Tor Vergata} University of Rome}
  \city{Rome}
   \country{Italy}
   }
   \email{luca.pepesciarria@gmail.com}

\author{Alessandro Straziota}
\affiliation{%
  \institution{\textit{Tor Vergata} University of Rome}
  \city{Rome}
   \country{Italy}
   }
    \email{alessandro.straziota@uniroma2.it}

\author{Matteo Stromieri}
\affiliation{%
    \institution{ \textit{Tor Vergata} University of Rome}
    \city{Rome}
    \country{Italy}
}
\email{matteo.stromieri@students.uniroma2.eu}

%%
%% The abstract is a short summary of the work to be presented in the
%% article.
\begin{abstract}
 In this work, we propose, analyze and empirically validate a lazy-update approach to  maintain accurate approximations  of  the $2$-hop neighborhoods of dynamic graphs resulting from sequences of edge insertions. 

We first show that under random input sequences, our algorithm exhibits an optimal trade-off between accuracy and insertion cost: it only performs $O(\frac{1}{\varepsilon})$ (amortized) updates per edge insertion, while the estimated size of any vertex's $2$-hop neighborhood is at most a factor $\varepsilon$ away from its true value in most cases, \emph{regardless} of the underlying graph topology and for any $\varepsilon > 0$. 

As a further theoretical contribution, we explore adversarial scenarios that can force our approach into a worst-case behavior at any given time $t$ of interest. 
We show that while worst-case input sequences do exist, a necessary condition for them to occur is that the \textit{girth} of the graph released up to time $t$ be at most $4$. 

Finally, we conduct extensive experiments on a collection of real, incremental social networks of different sizes, which typically have low girth. Empirical results are consistent with and typically better than our theoretical analysis anticipates. This further supports the robustness of our theoretical findings: forcing our algorithm into a worst-case behavior not only requires topologies characterized by a low girth, but also carefully crafted input sequences that are unlikely to occur in practice. 

Combined with standard sketching techniques, our lazy approach proves an effective and efficient tool to support key neighborhood queries on large, incremental graphs, including neighborhood size, Jaccard similarity between neighborhoods and, in general, functions of the union and/or intersection of $2$-hop neighborhoods.

\end{abstract}
\maketitle

%%% do not modify the following VLDB block %%
%%% VLDB block start %%%
\pagestyle{\vldbpagestyle}
\begingroup\small\noindent\raggedright\textbf{PVLDB Reference Format:}\\
\vldbauthors. \vldbtitle. PVLDB, \vldbvolume(\vldbissue): \vldbpages, \vldbyear.\\
\href{https://doi.org/\vldbdoi}{doi:\vldbdoi}
\endgroup
\begingroup
\renewcommand\thefootnote{}\footnote{\noindent
This work is licensed under the Creative Commons BY-NC-ND 4.0 International License. Visit \url{https://creativecommons.org/licenses/by-nc-nd/4.0/} to view a copy of this license. For any use beyond those covered by this license, obtain permission by emailing \href{mailto:info@vldb.org}{info@vldb.org}. Copyright is held by the owner/author(s). Publication rights licensed to the VLDB Endowment. \\
\raggedright Proceedings of the VLDB Endowment, Vol. \vldbvolume, No. \vldbissue\ %
ISSN 2150-8097. \\
\href{https://doi.org/\vldbdoi}{doi:\vldbdoi} \\
}\addtocounter{footnote}{-1}\endgroup
%%% VLDB block end %%%

%%% do not modify the following VLDB block %%
%%% VLDB block start %%%
\ifdefempty{\vldbavailabilityurl}{}{
\vspace{.3cm}
\begingroup\small\noindent\raggedright\textbf{PVLDB Artifact Availability:}\\
The source code, data, and/or other artifacts have been made available at \url{https://github.com/Gnumlab/graph_ball}.
%\url{\vldbavailabilityurl}.
\endgroup
}
%%% VLDB block end %%%

\section{Introduction}
In this paper, we consider the task of processing a possibly large, dynamic graph $G(V,E)$, incrementally provided as a stream of edge insertions, so that at any point of the stream it is possible to efficiently evaluate different queries that involve functions of the \textit{$h$-hop neighborhoods} of its vertices. For a vertex $v\in V$, its $h$-hop neighborhood is simply the \emph{set} of vertices that are within $h$ hops from $v$. In the remainder, $h$-hop neighborhoods are called \textit{$h$-balls} for brevity. As concrete examples of query types we consider, one might want to estimate the size of the $2$-ball at any given vertex, or the Jaccard similarity between the $2$-balls centered at any given two vertices, or other indices of a similar flavor that depend on the intersection or union between $1$-balls and/or $2$-balls, just to mention a few.

Neighborhood-based indices are common in key mining tasks, such as link prediction in social \cite{liben2003link} and biological networks \cite{wang2023assessment} or to describe statistical properties of large social graphs \cite{becchetti2008efficient}. For example, $2$-hop neighborhoods are important in social network analysis and similarity-based link prediction \cite{zhou2021experimental,zareie2020similarity,Sim-Nodes_Survey_2024}, while accurate approximations of $h$-balls' sizes are used to estimate key statistical properties of (very) large social networks \cite{boldi2011hyperanf,backstrom2012four}, or as link-based features in classifiers for Web spam detection \cite{becchetti2008link}. Further tasks that may explicitly involve or benefit from $2$-hop neighborhood queries are accurate estimates of centrality measures that are widely adopted in social network analysis \cite{freeman2002centrality,rochat2009closeness,bergamini2019computing} or mining of (usually large) bipartite networks that are frequent in certain applications \cite{sariyuce2018peeling,pavlopoulos2018bipartite}. We investigate the former example in detail in \Cref{apx:centrality}.

When the graph is static, an effective approach to this general task is to treat $h$-balls as subsets of the vertices of the graph, suitably represented using approximate summaries or sketches \cite{agarwal2013mergeable}. This line of attack has proved successful, for example in the efficient and scalable evaluation of important neighborhood-based queries on massive graphs that in part or mostly reside on secondary storage \cite{feigenbaum2005graph,mcgregor2014graph,becchetti2008efficient,boldi2011hyperanf}.
%and can only be accessed over consecutive, sequential passes

Nowadays, standard applications in social network analysis often entail dynamic scenarios in which  input graphs \textit{evolve over time}, under a sequence of  edge insertions and possibly deletions \cite{aggarwal2014evolutionary}. In many cases, such as co-authorship, citation or user-item association networks, the corresponding dynamic graphs are inherently or mostly \emph{incremental}, i.e., either edges and/or vertices are added but not deleted, or insertions account for the vast majority of updates \cite{leskovec2007graph,borner2004simultaneous}.\footnote{Of course, notable examples exist in which deletions are as important, such as some affiliation networks, or networks that describe on-line messaging behavior, where edges representing message exchanges become less and less important as they age.}

With respect to a static scenario, the dynamic case poses new and significant challenges even in the incremental setting, as soon as $h > 1$.\footnote{The case $h = 1$ is considerably simpler and it \textit{barely relates to graphs}: adding or removing one edge $(u, v)$ simply requires updating the $1$-balls of $u$ and $v$ accordingly, i.e., updating two corresponding set sketches by adding or removing one item. This has been the focus of extensive work in the recent past that we discuss in \Cref{subse:related}.} To see this, it may be useful to briefly sketch the cost of maintaining $1$- and $2$-balls exactly under a sequence of edge insertions, as we discuss in more detail in Section \ref{sec:detalgo}. 
When $h = 2$, each \texttt{Insert}($u, v$) operation entails (see Algorithm \ref{algo:naive} and Figure \ref{fig:basic-example}): i) updating the $2$-ball of $u$ to its union with the $1$-ball of $v$ and viceversa (what we call a \textit{heavy} update); ii) adding $v$ to the $2$-ball of \emph{each} neighbor of $u$ and viceversa (what we call a \textit{light} update). Both heavy and light updates can result in high computational costs per edge insertion: a heavy update can be expensive if at least one of the neighborhoods to merge is large; on the other hand, light updates are relatively inexpensive, but they can be numerous when large neighborhoods are involved, again resulting in a high overall cost per edge insertion. Unfortunately, $h$-balls can grow extremely fast with $h$ in many social networks, already as one switches from $h = 1$ to $h = 2$ \cite{becchetti2008link,backstrom2012four}. For the same reason, maintaining lossless representations of $2$-balls for each vertex of such networks might require considerable memory resources and might negatively impact the cost of serving neighborhood-based queries that involve moderately or highly central vertices.

To address the aforementioned issues for graphs that reside in main memory, one might want to trade some degree of accuracy for the following broad goals: 1) designing algorithms with low update costs, possibly $O(1)$ amortized per edge insertion; 2) minimizing memory footprint beyond what is needed to store the graph; 3) maintaining $1$- and $2$-balls using data structures that afford efficient, real-time computation of queries as the ones mentioned earlier with minimal memory footprint.

Heavy updates are natural and well-known candidates for efficient (albeit approximate) implementation using compact, sketch-based data structures \cite{gibbons2001estimating,broder2001completeness,broder2000identifying,agarwal2013mergeable,trevisan/646978.711822}. However, sketches alone are of no avail in handling light updates, whose sheer potential number requires a novel approach.
The literature on efficient data structures that handle insertions and often deletions over dynamic graphs is rich. However, efficient solutions to implement neighborhood-based queries on dynamic edge streams are only known for $1$-balls \cite{BSS20,MROS,VOS,CGPS24}, nor do approaches devised for other dynamic problems adapt to our setting in any obvious way, something we elaborate more upon in Section \ref{subse:related}.

\subsection{Our Contribution}
In this paper, we propose an approach that trades some degree of accuracy for a substantial improvement in the average number of light updates. In a nutshell, upon an edge insertion, our algorithm performs the (two) corresponding heavy updates, but in general only a subset of the required light updates, according to a scheme that combines a threshold-based mechanism and a randomized, batch-update policy. Hence, for every vertex $u$, we only keep an approximation (a subset to be specific) of $u$'s $2$-ball. If $1$- and $2$-balls are represented with suitable data sketches, our approach affords constant average update cost per edge insertion.\footnote{The particular sketch used depends on the neighborhood queries we want to be able to serve. When sketches are used, the cost of merging two neighborhoods corresponds to the cost of combining the corresponding sketches, which is typically a constant that depends on the desired approximation guarantees. For example, if we are interested in the Jaccard similarity between pairs of $1$- and/or $2$-balls, this cost will be proportional to the (constant) number of minhash values we use to represent each neighborhood.} While the behavior and accuracy guarantees of most sketching techniques are well understood, the estimation error induced by lazy updates can be arbitrarily high in some cases. The main focus of this paper is on the latter aspect, which is absent in the static case but critical in the dynamic setting. Accordingly, we assume lossless representations of $1$-balls and approximate $2$-balls in our theoretical analyses in \Cref{sec:detalgo,subse:rand_perm,sec:gammaok}, while we use  standard sketching techniques to represent $1$- and $2$-balls in the actual implementations of the algorithms and baselines we consider in the experimental analysis discussed in \Cref{sec:exp}.

\paragraph{Almost-optimal performance on random sequences.} We prove in Section \ref{subse:rand_perm} that even a simplified, deterministic variant of our lazy-update \Cref{alg:det_thresh} achieves asymptotically optimal expected performance when the  sequence of edge insertions is a random, uniform permutation over an \textit{arbitrary} set of edges. In other words, our lazy approach is robust to adversarial topologies as long as the edge sequence follows a random order.
Formally, we prove that, for any desired $0 < \varepsilon < 1$, our algorithm only performs $O(\frac{1}{\varepsilon})$ (amortized) updates per edge insertion, while at any time $t$ and for every vertex $v$, the estimated size of $v$'s $2$-ball is, in expectation,  at most a factor $\varepsilon$ away from its true value. We further prove that this approximation result holds with a probability that exponentially increases with the true size of the $2$-ball itself (\Cref{thm:random_seq_quality}). 
Thanks to this analysis in concentration,  our results can be extended to other functions of $2$-balls, including union, intersection and Jaccard similarity (see \Cref{cor:jacc} for this less obvious case).

\smallskip
As positive as this result may sound, it begs the following questions from a careful reader: 1) Are the results above robust to adversarial sequences? 2) Is a performance analysis under random sequences representative of practical scenarios? 3) More generally, does our lazy scheme offer significant practical advantages?

\paragraph{Performance analysis on adversarial inputs.} While our results for random sequences are optimal regardless of the underlying graph's topology, one might wonder about the ability of an adversary to design \textit{worst-case, adaptive sequences} that force our approach to behave poorly and, in this case, whether any conditions on the graph topology are \textit{necessary} for this to happen. We investigate this issues in Section \ref{sec:gammaok}, where we first show that it is possible to design worst-case sequences of edge insertions that force our algorithm to perform arbitrarily worse than the random setting (\Cref{thm:lower}). However, as a further contribution, we also prove that worst-case input sequences exist \textit{only if} the \textit{girth} \cite{diestel2024graph} of the final graph is at most $4$.
More precisely, we show that a randomized, special case of \Cref{alg:det_thresh} achieves asymptotically optimal performance on a class of graphs that contains all graphs with girth at least $5$, \footnote{The  class is in fact more general since it also includes graphs with  a ``bounded'' number of cycles of length at most 4. See \Cref{def:gammaok}, for a formal definition of this class.} even when the input sequence is chosen by an adaptive adversary. 

\smallskip

\paragraph{Experimental analysis.} 
An analysis under random permutation sequences is relatively common in the literature on dynamic edge streams and data streams \cite{buriol2006counting,kapralov2014approximating,peng2018estimating,Hanauer22DynamicSurvey}. Yet, one might reasonably wonder about its practical significance for the task considered in this paper. We address these issues in Section \ref{sec:exp}, where we conduct experiments on small, medium and large-sized, incremental graphs (whose main properties are summarized in Table \ref{tab:summary_dynamic_dataset}). At least on the diverse sample of real networks we consider, experimental results on the estimation of key queries such as size and Jaccard similarity are consistent with the theoretical findings from Section \ref{subse:rand_perm}, while results on running times highlight the significance of our approach in practice: i) light updates crucially affect computational costs in practical scenarios; ii) when combined with sketches, our approach strikes an excellent balance between loss in accuracy and computational efficiency. In particular, its accuracy is comparable or only slightly worse than the baseline that performs all necessary light updates, while it achieves speed-ups that increase with network size and can be orders of magnitude with respect to the baseline on large or very large social networks (Tables \ref{tab:size-time} and \ref{tab:size-time-friendster}).
We finally note that our datasets are samples of real social networks. As such, they have relatively large local and global clustering coefficients\footnote{At least the undirected ones.} and thus low girth. Hence, our experimental analysis further supports the robustness of our theoretical findings: forcing our algorithm(s) into a worst-case behavior not only requires topologies characterized by a low girth, but also carefully crafted input sequences that are unlikely to occur in practice.

\paragraph{Use cases from real networks.} 
While neighborhood-based queries are extremely important in social network analysis, the main focus of this paper is on crucial, new challenges that arise in dynamic settings and on possible strategies to address them. For this reason, we investigate basic neighborhood queries without specific applications in mind. 
As simple as it may appear however, the incremental setting we consider provides a feasible framework to tackle important mining tasks over large, dynamic   networks. As an example, we consider the problem of maintaining the \textit{harmonic centrality}-based ranking of the vertices of an incremental graph \cite{rochat2009closeness, boldi2014axioms}, a key mining task in social network analysis \cite{boldi2014axioms,freeman2002centrality}. Exact solutions for this problem necessarily incur high computational costs in some cases,\footnote{Even in the static setting \cite{bergamini2019computing}.} while the study of dynamic strategies with update costs of practical interest is still an open area of research \cite{HanauerHS22}. In \Cref{apx:centrality}, we experimentally show that our approach is a very good fit for this task, affording continuous, accurate tracking of top-$k$ ranking vertices, thus highlighting its effectiveness for a key downstream task that would otherwise be hardly feasible on large, incremental graphs.

Many real networks are inherently bipartite, describing interactions between entities of two different types. Examples include many biological networks  \cite{pavlopoulos2018bipartite}, social networks \cite{LATAPY200831}, user-URL or user-item interactions \cite{leskovec2009meme} to name a few. Moreover, many of them are also inherently or mostly incremental in nature.\footnote{Edge/vertex deletions are rare or absent.} In such networks, the prediction of future interactions is a key mining task, for which (Jaccard) similarity between $2$-hop neighborhoods can prove a valuable index for prediction purposes. For example, in author-paper networks, the set $L_2(v)$ of vertices at distance exactly $2$ from an author $v$ is the set of $v$'s co-authors. Hence, for any two authors $u$ and $v$ with no currently co-authored paper, the Jaccard similarity between  $L_2(u)$ and $L_2(v)$ can be used to estimate the probability of $u$ and $v$ co-authoring a paper in the next future.\footnote{Our approach also allows estimation of the Jaccard similarity between $\lset_2(u)$ and $\lset_2(v)$.} 

\paragraph{Final remarks.}
We stress that although we present them for the undirected case for ease of presentation and for the sake of simplicity, our algorithms apply to directed graphs in general,\footnote{Of course, in this case we have directed $h$-balls, i.e., sets of a vertices that can be reached in at most $h$ hops from a given vertex, or from which it is possible to reach the vertex under consideration in at most $h$ hops.} while our analysis extends to the directed case with minor modifications. Moreover, our algorithms seamlessly address incremental settings in which new vertices may connect to the current graph, the latter being for example the typical scenario in co-authorship networks \cite{borner2004simultaneous}.

\subsection{Further related work}\label{subse:related}
Efficient data structures for queries that involve $h$-balls of a dynamic graph turn out to be useful in different network applications, as we  mentioned earlier \cite{becchetti2008link,zareie2020similarity,Sim-Nodes_Survey_2024}.

\iffalse
Efficient data structures for queries that involve $h$-balls of a dynamic graph turn out to be useful in different network applications. Besides those we  mentioned earlier   \cite{becchetti2008link,zareie2020similarity,Sim-Nodes_Survey_2024}, we cite here the work \cite{cavallo20222}, where the notion of \textit{2-hop Neighbor Class Similarity} (2NCS) is proposed: this is  a new quantitative graph structural property that correlates with \textit{Graph Neural Networks} (GNN) \cite{scarselli2008graph,wu2022graph} performance more strongly and consistently than alternative metrics. 2NCS considers two-hop neighborhoods as a theoretically derived consequence of the two-step label propagation process governing GNN’s training-inference process.
\fi

%\item \textbf{Efficient solutions for \textit{all our queries} on  1-Balls on dynamic graphs.} 

As remarked in the introduction, efficient solutions for Jaccard similarity queries on $1$-balls   have been proposed for different dynamic graph models: all of them share the use of suitable data sketches to manage insertion and deletion of elements from sets. In particular,  \cite{BSS20,MROS,VOS} proposes and compares different  approaches that work in  the fully-dynamic streaming model, while an efficient solution, based on a buffered version of the $k$-min-hashing scheme is proposed in \cite{CGPS24}. This works in the fully-dynamic streaming model and allows recovery actions when certain ``worst-case'' edge deletion events occur. A further algorithm is presented in \cite{zhang2022effective}, where \textit{bottom-$k$ sketches} \cite{cohen2007summarizing} are used to perform dynamic graph clustering based on Jaccard similarity among vertices' neighborhoods. We remark that 
none of these previous approaches include ideas or tools that can be adapted to efficiently manage the $2$-ball update-operations we need to implement in this work.

As for other queries that might be "related" or "useful" in our setting, a considerable amount of work on data structures that support edge insertions and deletions exists for several queries,  such as connectivity or reachability, (exact or approximate) distances, minimum spanning tree, (approximate) \textit{betweenness centrality}, and so on. We refer the reader to \cite{HanauerHS22} for a nice survey on experimental and theoretical results on the topic. To the best of our knowledge however, none of these approaches can be obviously adapted to handle the types of queries we consider in this work. For example, a natural idea would be using an incremental data structure to dynamically maintain the first $h$ levels of a BFS tree, such as \cite{EvenS81,RodittyZ11}, that achieve $O(h)$ amortized update time. However, let alone effectiveness in efficiently serving queries as the ones we consider here, the data structure uses $\Omega(n)$ space per BFS. This is prohibitive in our setting, where we would need to instantiate one such data structure for each vertex, with total space $\Omega(n^2)$. Moreover, since in a degree-$\Delta$ graph $\Theta(\Delta)$ BFS trees can change following a single edge insertion, the corresponding amortized time per edge insertion could be as high as $\Theta(\Delta)$, which is basically the same cost of the baseline solution we discuss at the beginning of Section \ref{sec:detalgo}.

\section{Lazy-Update Algorithms} \label{sec:detalgo}

After giving some preliminaries we will use through all this paper, in \Cref{ssec:algos} we describe the lazy-update algorithmic scheme, while in \Cref{ssec:detalgo-time-wc}, we provide a general bound on its amortized update cost that holds for  arbitrary sequences of edge insertions.

\subsubsection*{Preliminaries and notations}
The dynamic (incremental) graph model we study can be defined as a sequence
    $\dynG = \{ G^{(0)}(V,E^{(0)}), \ldots, $ $ G^{(t)}(V,E^{(t)}),  \ldots G^{(T)}(V,E^{(T)}) \}$,   where: (i) the set of vertices $V = \{1, \ldots , n \}$ is fixed, (ii) $T \leq \binom{n}{2}$ is the final graph,  while (iii)  $E^{(t)}$ is the subset of edges at time $t$. Note that this  changes in every time step $t \geq 1$, as a new edge $e^{(t)}$ is inserted, so that $E^{(t+1)} = E^{(t)} \cup \{e^{(t)}\} $. 
We remark that  our analysis and all our results can be easily adapted to a more general model that includes any combination of the following variants: (i)  growing vertex sets, (ii)  multiple insertions of the same edge, and (iii) directed edges (thus yielding directed graphs).  However,  the corresponding  adaptations of our analysis  would require  significantly heavier notation and some technicalities that we decided to avoid for the sake of clarity.

Our goal is to design algorithms that, at every time step $t \geq 1$, are able to efficiently compute queries over the current $2$-balls of $G^{(t)}$. As mentioned in the introduction, our focus is on queries that are typical in graph mining such as: (i) given a vertex $u$, estimate the size of $\ball_2(u)$, and (ii) given two vertices $u,v \in V$, estimate the Jaccard similarity of the corresponding $2$-balls:
\[ 
    \jacc(\ball_2(u),\ball_2(v)) \ = \ \frac{|\ball_2(u) \cap \ball_2(u) |}{|\ball_2(u) \cup \ball_2(u) |}  \, . 
\]
Both the  theoretical  and experimental analysis of  our   lazy-update algorithms  consider the following key performance  measures: the \textit{amortized update time} per edge insertion and the \textit{approximation ratio} of our algorithms on the quantities $|\ball_2(u)|$ and $\jacc(\ball_2(u),\ball_2(v))$, for any choice of the input vertices. Intuitively, the amortized update time is the average time it takes to process a new edge, a more formal definition is deferred to \Cref{sec:detalgo}, after a detailed description of the algorithms  we consider.
    
We next summarize notation that is used in the remainder of the paper. For a vertex $v \in V$ of a graph $G(V,E)$, we define:

\begin{description} 
    \item[$\neigh(v)$:] the set of neighborhoods of the vertex $v$.
    \item[$\deg_v$:] the degree of $v$. Notice that $\deg_v = \vert \neigh(v) \vert$;
    \item[$\lset_h(v)$:] set of vertices at distance exactly $h$ from $v$;
    \item[$\ball_h(v)$:] set of vertices at distance at most $h$ from $v$. 
\end{description}

The reader may have noticed that, in our notation above, the term $t$ does not appear: this is due to the fact that our analysis holds at any (arbitrarily fixed) time step, which is always clear from context.

\subsection{Algorithm description} \label{ssec:algos}

%\subsection{A threshold-based deterministic algorithm}\label{subse:threshold}
Consider the addition of a new edge $(u,v)$ to $G$. Clearly, the only $2$-balls that are affected are those centered at $u$, $v$, and at every vertex $w \in \neigh(u) \cup \neigh(v)$. A \emph{baseline} strategy, given as \Cref{algo:naive} for the sake of reference, tracks changes exactly and thus updates all $2$-balls that are affected by an edge insertion.

% \begin{algorithm}[h!]
% \SetAlgoLined
% \DontPrintSemicolon
% %\KwData{Undirected Graph $G = (V, E)$}
% \SetKwFunction{FMain}{Insert}
% \SetKwProg{Fn}{Function}{:}{end}
% \Fn{\FMain{$u, v$}}{
%     \ForEach{$x \in \neigh(u) \cup \{ u \}$}{
%         $\ball_2(v) \gets \ball_2(v) \cup \{x\}$\;
%         $\ball_2(x) \gets \ball_2(x) \cup \{v\}$\;
%     }
%     \ForEach{$x \in \neigh(v) \cup \{v\}$}{
%         $\ball_2(u) \gets \ball_2(u) \cup \{x\}$\;
%         $\ball_2(x) \gets \ball_2(x) \cup \{u\}$\;
%     }
% }
% \caption{Baseline algorithm.}
% \label{algo:naive}
% \end{algorithm}

\begin{algorithm}[h!]
\SetAlgoLined
\DontPrintSemicolon
\SetKwFunction{FMain}{Insert}
\SetKwProg{Fn}{Function}{:}{end}
\Fn{\FMain{$u, v$}}{
    % \For{$x \in \{u,v\}$}{
    %     let $y \in \{u,v\} \setminus \{x\}$\;
    %     $\ball_2(y) \gets \ball_2(y) \cup \neigh(x) \cup \{x\}$\;
    %     \ForEach{$z \in \neigh(u)$}{
    %         $\ball_2(z) \gets \ball_2(z) \cup \{y\}$\;
    %     }
    % }
    $\ball_2(u) \gets \ball_2(u) \cup \ball_1(v)$\;\label{alg:naive:line:heavy_up_1}
    \ForEach{$w \in \neigh(u) \setminus \{v\}$}{
        $\ball_2(w) \gets \ball_2(w) \cup \{v\}$\; \label{alg:naive:line:light_up_1}
    }
    $\ball_2(v) \gets \ball_2(v) \cup \ball_1(u)$\;\label{alg:naive:line:heavy_up_2}
    \ForEach{$w \in \neigh(v) \setminus \{u\}$}{
        $\ball_2(w) \gets \ball_2(w) \cup \{u\}$\; \label{alg:naive:line:light_up_2}
    }
}
\caption{Baseline algorithm.}
\label{algo:naive}
\end{algorithm}

The magnitude of the changes (and the associated computational costs) induced by \texttt{Insert}$(u, v)$ vary. In particular, $\ball_2(u)$ can change significantly, as all vertices in $\ball_1(v)$ will be included in $\ball_2(u)$ (we refer to this as a \emph{heavy} update, see \Cref{algo:naive} at \Cref{alg:naive:line:heavy_up_1}). Instead, for any vertex $w \in \neigh(u)\setminus\{v\}$, $\ball_2(w)$ will grow by at most one element, namely $v$ (this is referred to as a \emph{light} update, see \Cref{algo:naive} at \Cref{alg:naive:line:light_up_1}). 
Symmetrically, the same holds for $v$ and for every $w \in \neigh(v) \setminus \{u\}$ (\Cref{alg:naive:line:heavy_up_2,alg:naive:line:light_up_2}). 

A key observation at this point is that, while heavy updates can be addressed using (possibly, approximate) data structures that allow efficient merging of $1$- and $2$-balls, this line of attack fails with light updates, whose cost derives from their potential number, which can be large in many real cases, as we noted in the introduction.

\begin{figure}[ht]
    \centering
    \includegraphics[width=0.45\linewidth]{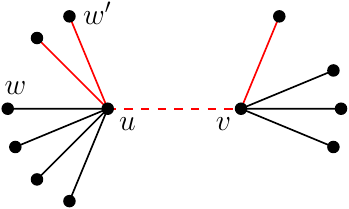}
    \caption{Example of insertion of a new edge $(u,v)$. The algorithm merges the $1$-ball of $v$ with the $2$-ball of $u$ (heavy update), while it does not immediately add vertex $v$ to the $2$-ball of vertex $w$ or any other of $u$'s neighbors. 
    %This is not the only vertex $w$ is not aware of (for instance $w'$ is another vertex %belonging to the 2-hop ball of $w$ that $w$ is not aware of).
    }
    \label{fig:basic-example}
\end{figure}

A first idea to reduce the average number of updates per edge insertion is to perform heavy updates immediately, instead processing light updates in batches that are performed occasionally. More precisely, when a new edge $(u,v)$ arrives, it is initially marked as a \emph{red edge}. Whenever the number of red edges incident to a vertex $u$ exceeds a certain threshold, all the corresponding light updates are processed, and the state of red edges is updated to \emph{black}. See \Cref{fig:basic-example} for an example. 

The idea behind the threshold-based approach is to maintain a balance between the number of black and red edges for every vertex. While useful when edge insertions appear in a random order, this approach may fail when red edges considerably expand the original size of the $2$-ball of some vertex $u$. In order to mitigate this problem, our algorithm  uses a second ingredient: upon each edge insertion $(u,v)$, the algorithm selects $k$ vertices from $\neigh(u)$ and $k$ from $\neigh(v)$ uniformly at random and performs a batch of light updates for the selected vertices, even if the threshold has not been reached yet.

\iffalse
Notice that, if the number of red edges did not reach the threshold yet, there might be some vertex $w \in \neigh(u) \setminus \{v\}$ that is not aware of all the vertices contained in its own $2$-ball. In order to mitigate this, our algorithm will use another ingredient. At each edge insertion $(u,v)$, the algorithm selects uniformly at random $k$ vertices from $\neigh(u)$ and $k$ from $\neigh(v)$, and performs the batch of light updates for the selected vertices, even if the threshold has not yet been reached.
\fi

These ideas are formalized in \Cref{alg:det_thresh}. For each vertex $v$, our algorithm maintains two sets $\apxball_1(v)$ and $\apxball_2(v)$, as well as the \emph{black degree} $\bd_v$ and \emph{red degree} $\rd_v$ of $v$.
Our algorithm guarantees that $\apxball_1(v)$ is exactly $\ball_1(v)$, while $\apxball_2(v)$ is in general a subset of $\ball_2(v)$. The algorithm uses two global parameters, namely a \textit{threshold} $\varphi \in [0,1]$, and an integer $k$. The role of the parameter $\varphi$ can be understood as follows: when $\varphi$ is set to $0$, the algorithm performs all heavy and light updates for every edge insertion, ensuring that $\apxball_2(v)$ always matches $\ball_2(v)$. As $\varphi$ increases, the update function becomes lazier: light updates are not always executed, and $\apxball_2(v)$ is typically a proper subset of $\ball_2(v)$. For instance, when $\varphi = 1$, light updates are performed in batches every time the degree of a vertex doubles. Parameter $k$ specifies the number of neighbors of $v$ that are randomly selected for update of their $2$-balls whenever an edge insertion involving $v$ occurs. This mechanism corresponds to \Cref{line:random_selection,line:random_selection_for,line:random_selection_for_inside} of \Cref{alg:det_thresh}. 

We call \lazyscheme$(\varphi,k)$ the algorithm that runs \Cref{alg:init} on an initial graph $G^{(0)}$ and then processes a sequence $S$ of edge insertions by running \Cref{alg:det_thresh} on each edge of $S$.  

\begin{algorithm}[h]
\SetAlgoLined
\DontPrintSemicolon
\KwData{An undirected graph $G=(V,E)$, a threshold parameter $0 \leq \varphi \leq 1$, and an integer $k \geq 0$.}
set $\varphi$ and $k$ as global parameters\;
\ForEach{vertex $u \in V$}{
    $\delta_u \gets 0$\;
    $\Delta_u \gets \deg_u$\;
    $\apxball_1(u) \gets \ball_1(u)$\;
    $\apxball_2(u) \gets \ball_2(u)$\;
}
\caption{\texttt{Init} operation}\label{alg:init}
\end{algorithm}

%Due to the nature of this algorithm, which triggers batch updates once a certain threshold %is surpassed, we have named it \emph{Threshold-Batching Update}.

\iffalse
When a new edge is added to the graph, the algorithm updates the above information as detailed in the pseudo-code given in \Cref{alg:det_thresh}. The role of the parameter $\varphi$ can be understood as follows: when $\varphi$ is set to $0$, the algorithm performs all the heavy and light updates for every edge insertion, ensuring that $\apxball_2(v)$ always matches the current ball $\ball_2(v)$. As $\varphi$ increases, the update function becomes lazier: light updates are not always executed, and $\apxball_2(v)$ may become a strict subset of $\ball_2(v)$. For instance, when $\varphi = 1$, light updates are performed in batches every time the degree of a vertex doubles. \Cref{line:random_selection,line:random_selection_for,line:random_selection_for_inside} specify the random selection explained above.

Due to the nature of this algorithm, which triggers batch updates once a certain threshold is surpassed, we have named it \emph{Threshold-Batching Update}.
\fi

\begin{algorithm}[h]
\SetAlgoLined
\DontPrintSemicolon
%\KwData{An undirected graph $G=(V,E)$, a threshold factor $\varphi$.}
\SetKwFunction{FMain}{Insert}
\SetKwProg{Fn}{Function}{:}{end}
\Fn{\FMain{$(u, v)$}}{
    \For{$x \in \{u,v\}$}{
        let $y \in \{u,v\} \setminus \{x\}$\;
        $\apxball_1(x) \gets \apxball_1(x) \cup \{y\}$\; \label{line:simple_union}
        \tcp{heavy update}
        $\apxball_2(x) \gets \apxball_2(x) \cup \apxball_1(y)$\; \label{line:heavy_update}
        $\delta_x \gets \delta_x + 1$\;
    
        \uIf{$\delta_x \geq \varphi \cdot \Delta_x$}{ \label{line:threshold_check}
            $\Delta_x \gets \Delta_x + \delta_x$\;
            $\delta_x \gets 0$\;
            \ForEach{$z \in \neigh(x)$}{ \label{line:propagate}
                \tcp{batch of light updates}
                $\apxball_2(z) \gets \apxball_2(z) \cup \apxball_1(x)$\; \label{line:light_updates_for}
            }
      }
      \Else {
        select $k$ vertices $w_1, \dots, w_k \in \neigh(x)$ u.a.r.\;\label{line:random_selection}
        \For{$i = 1, \dots, k$}{ \label{line:random_selection_for}
                \tcp{batch of light updates}
                $\apxball_2(w_i) \gets \apxball_2(w_i) \cup \apxball_1(x)$\; \label{line:random_selection_for_inside}
            } 
      }
    }
}

\caption{ \textsc{Insert} }\label{alg:det_thresh}
\end{algorithm}
\paragraph{A note on neighborhood representation.}
As we mentioned in the introduction, we treat  $\apxball_1(v)$ and $\apxball_2(v)$ as sets of vertices in this section and in Section \ref{sec:gammaok}. We remark that this only serves the purpose of analyzing the error introduced by our lazy update policies: lossless representations of $1$- and $2$-balls may be unfeasible for medium or large graphs and compact data sketches are typically used to represent them in such cases. The choice of the actual sketch strongly depends on the type of query (or queries) one wants to support, such as $1$- or $2$-ball sizes \cite{flajolet1985probabilistic,boldi2011hyperanf} or Jaccard similarity between $2$-balls \cite{broder2000identifying,cohen2007summarizing,becchetti2008efficient}. 
All sketches used for typical neighborhood queries are well-understood and come with strong accuracy guarantees. Moreover, they allow to perform the union of $1$- and $2$-balls we are interested in time proportional to the sketch size, which is independent of the sizes of the balls to merge \cite{agarwal2013mergeable}. 

\subsection{Cost analysis for arbitrary sequences} \label{ssec:detalgo-time-wc}
Consistently to what we remarked above, our cost analysis focuses on the number of 
\emph{set-union} operations: This performance measure in fact dominates
the computational cost of \Cref{alg:det_thresh}.  More in detail, given any sequence $S$ of edge insertions, starting from an initial graph $G^{(0)}$,  we denote by $\cost(S)$ the overall number of union operations performed in \Cref{line:simple_union,line:heavy_update,line:light_updates_for,line:random_selection_for_inside} of \Cref{alg:det_thresh} on the input sequence $S$.

We observe that a trivial upper bound to $\cost(S)$ is $O(\Delta |S|)$, since each insertion can cost $O(\Delta)$ union operations where $\Delta$ is the maximum degree of the current graph. However, this trivial argument turns out to be too pessimistic: in what follows,  we  provide a  more refined analysis of the amortized cost\footnote{The \emph{amortized analysis} is a well-known method originally introduced in \cite{Tarjan_amortized} that allows to compute tight bounds on the cost of a \textit{sequence} of operations, rather than the worst-case cost of an individual operation. In more detail,  we average the cost of a \emph{worst case} sequence of operations to obtain a more meaningful cost per operation.} per edge insertion. We say that an algorithm has \emph{amortized cost} $\hat{c}$ per edge insertion if, for any sequence $S$ of edge insertions, we have $\cost(S) \le \hat{c} |S|$.

\begin{lemma}
    \label{lm:amortized_det_alg}
    Given any initial graph $G^{(0)}$ and any sequence $S$ of edge insertions, the amortized update cost of \Cref{alg:det_thresh} is $O(\frac{1}{\varphi}+k)$ per edge insertion.
\end{lemma}
\begin{proof}
    Let us first consider the case $k=0$, i.e., when the random selection and the consequent instructions in \Cref{line:random_selection,line:random_selection_for,line:random_selection_for_inside} are never performed.  Our amortized analysis makes use of the \textit{accounting method} \cite{Tarjan_amortized}. The idea is  paying  the cost of any batch of light updates by charging it to previous edge insertions. More precisely, we assign \emph{credits} to each edge insertion that we will use to pay the cost of subsequent batches of light updates. Formally, the \emph{amortized} cost of an edge insertion is defined as the \emph{actual} cost of the operation, plus the credits we assign to it, minus the credits (accumulated from previous operations) we spend for it. We need to carefully define such credits in order to guarantee that the sum of the amortized costs is an upper bound to the sum of the actual costs, i.e. we always have enough credits to pay for costly batch light updates.
    
    We proceed as follows. When we insert the edge $(u,v)$, we put $2/\varphi$ credits on $u$ and $2/\varphi$ credits on $v$. Now we bound the actual and amortized cost of each insertion. 
    
    First, consider an edge insertion $(u,v)$ that does not trigger a batch of light updates. Its actual cost is 4 union operations (those in \Cref{line:simple_union,line:heavy_update}, 2 for each endpoint of $(u,v)$). Then its amortized cost is upper-bounded by $4+4/\varphi=O(1/\varphi)$. Now consider the case in which the insertion causes a batch of light updates for $u$, or $v$, or both. We show that the credits accumulated by previous insertions are sufficient to pay for its cost. To see this, consider a batch of light updates involving vertex $x \in \{u,v\}$. And let $\bd_x$ and $\rd_x$ be the current black and red degrees of $x$ at that time (just before \Cref{line:threshold_check} is evaluated). It is clear that for vertex $x$ we have accumulated  $\rd_x \cdot 2/\varphi$ credits that now we use to pay for the cost of \Cref{line:propagate,line:light_updates_for}. This cost equals to $\deg_x$ union operations, where $\deg_x$ is the current degree of vertex $x$. Since the batch of light updates has just been triggered, we have that $\rd_x \ge \varphi \bd_x$, and hence we have at least $\rd_x \cdot 2/\varphi \ge \varphi \bd_x \cdot 2/\varphi=2 \bd_x$ credits to pay for the $\deg_x=\bd_x+\rd_x=\bd_x+\varphi\bd_x \le 2 \bd_x$ union operations. This concludes the proof. 

    Finally, to obtain the claim when $k>0$, we notice that, in this case, every edge insertion causes $O(k)$ additional union operations.
\end{proof}

%\subsection{Approximation analysis over random  sequences of arbitrary graphs}
\section{Random edge sequences}\label{subse:rand_perm}
In this section, we analyze the accuracy of our lazy-update algorithm(s) over an arbitrary dynamic graph, whose edges are given in input as a uniformly sampled, random permutation over its edge set.
Dynamic graphs resulting from random sequences of edge insertions have been an effective tool to provide theoretical insights that have often proved robust to empirical validation in various dynamic scenarios \cite{monemizadeh2017testable,kapralov2014approximating,peng2018estimating,mcgregor2014graph,chakrabarti2008robust}.
In more detail, assume $G = (V, E)$, with $|E| = t$, is the graph observed up to some time $t$ of interest. Following \cite{monemizadeh2017testable,peng2018estimating}, we assume that the sequence of edges up to time $t$ is chosen uniformly at random from the set of all permutations over $E$.\footnote{It should be noted that this includes the general case in which $t$ is any intermediate point of a longer stream that possibly extends well beyond $t$. In this case, it is well-known and easy to see that, conditioned on the subset $E$ of the edges released up to time $t$, their sequence is just a permutation of $E$.} The following fact is an immediate consequence of well-known and intuitive properties of random permutations. We state it informally for the sake of completeness, avoiding any further, unnecessary notation.
\begin{fact}\label{fa:perm}
    Consider a dynamic graph $G = (V, E)$, whose edges are observed sequentially according to a permutation over $E$ chosen uniformly at random. Then, for every $E'\subseteq E$, the sequence in which edges in $E'$ are observed is itself a uniformly chosen, random permutation over $E'$.
\end{fact}

In the remainder, for an arbitrary vertex $v$, we analyze how well the output $\apxball_2(v)$ of \Cref{alg:det_thresh} approximates $\ball_2(v)$ at any round $t$ in terms of its \emph{coverage}:

\begin{definition}\label{def:coverage}
    We say that the output  $\apxball_2(v)$  of \lazyscheme$(\varphi,k)$ is a $(1-\varepsilon)$-\textit{covering} of $\ball_2(v)$ if the following holds: i) $\apxball_2(v) \subseteq \ball_2(v)$; ii) $\Expec{}{\vert \apxball_2(v) \vert} \geq (1-\varepsilon) \vert \ball_2(v) \vert$, where expectation is taken over the randomness of the algorithm and/or the input sequence. When the algorithm produces a $(1-\varepsilon)$-covering $\apxball_2(v)$ of $\ball_2(v)$ for every $v$, we say it has \emph{approximation ratio} $\frac{1}{(1-\varepsilon)}$.
\end{definition}
Our main result in this section is formalized in the following 

\begin{theorem}\label{thm:random_seq_quality}
    Let  $\varepsilon \in (0,1)$,  and fix  parameters $k = 0$ and $\varphi = \frac{\varepsilon}{1-\varepsilon}$. Consider any  graph $G(V,E)$ submitted as  a  uniform  random permutation of its edge set $E$ to \lazyscheme$(\varphi,k)$. Then, at every time step $t\le |E|$, the algorithm has approximation ratio $\frac{1}{1-\varepsilon}$. Moreover, for every $\alpha > 0$ and every vertex $v  \in V$, we have:
    \begin{equation}\label{eq:size_prob_bound}
        \Prob{}{|\apxball_2(v)| < \frac{1 - \alpha}{1 + \varphi}|L_2(v)|}\le e^{-\frac{2\alpha^2|L_2(v)|}{(1 + \varphi)^2}}.
    \end{equation}
\end{theorem}
\begin{proof}
Fix a vertex $v \in V$ and a round $t \geq 1$. In the remainder of this proof, all quantities are taken at time $t$. We are interested in how close $|\apxball_2(v)|$ is to $|\ball_2(v)|$. To begin, we note that the following relationship holds deterministically:
\begin{equation}\label{eq:apxball_det}
    |\apxball_2(v)| = 1 + |L_1(v)| + |\apxball_2(v)\cap L_2(v)|,
\end{equation}
where the only random variable on the right hand side is the last term. 
\begin{figure}[h!]
    \centering
    \includegraphics[width=0.7\linewidth]{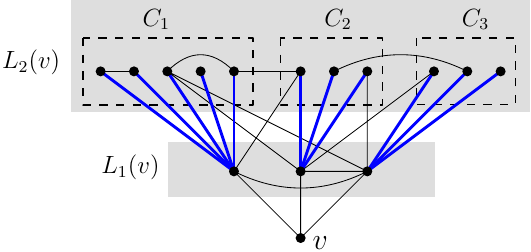}
    \caption{Example of a partition of $L_2(v)$ into three sets $C_1, C_2, C_3$. Edges connecting vertices $w \in L_2(v)$ to their respective partitions are thicker.}
    \label{fig:L2_partition}
\end{figure}
We next define a partition $\mathcal{C} = \{C_{u}: u\in L_1(v)\}$ of $L_2(v)$ as follows: for each $w \in L_2(v)$, we choose a vertex $u \in L_1(v) \cap \neigh(w)$ and assign $w$ to $C_u$. This way, each vertex $w\in L_2(v)$ is associated to exactly \emph{one} edge connecting one vertex in $L_1(v)$ to $w$ (see \Cref{fig:L2_partition}, where the edges in question are thick in the picture). Let $E_v$ denote the set of such edges and note that i) $E_v$ is a subset of the edges connecting vertices in $L_1(v)$ to those in $L_2(v)$, ii) $|E_v| = |L_2(v)|$ by definition and iii) $|C_u|\le \deg_u - 1$ for every $u\in L_1(v)$, given that $C_u$ contains a subset of $u$'s neighbors and $(v, u)$ is always present. Moreover, \Cref{alg:det_thresh} guarantees that $|\apxball_2(v)\cap L_2(v)|$ is at least the number of edges in $E_v$ that are black. 
These considerations allow us to conclude that
\[
    |\apxball_2(v)\cap L_2(v)| \ge |\{e\in E_v:\text{ $e$ is black}\}|.
\]
A key observation at this point is that \Cref{alg:det_thresh} implies that for every $x\in V$, $\rd_x\le\left\lfloor\frac{\varphi}{1 + \varphi}\deg_x\right\rfloor$. As a consequence, if some $e = (u, w)\in E_v$ was not among the last $\left\lfloor\frac{\varphi}{1 + \varphi}\deg_u\right\rfloor$ edges incident in $u$ that were released within time $t$, it is necessarily black. For $e = (u, w)\in E_v$, with $u\in L_1(v)$ and $w\in L_2(v)$, let $X_e = 1$ if $e$ was among the first $\deg_u - \left\lfloor\frac{\varphi}{1 + \varphi}\deg_u\right\rfloor$ edges incident in $u$ that were released up to time $t$ and let $X_e = 0$ otherwise. Following the argument above, the event $( X_e = 1 )$ implies the event $\text{"$e$ is black"}$, whence:
\begin{equation}\label{eq:apx_balls}
    |\apxball_2(v)\cap L_2(v)| \ge |\{e\in E_v:\text{ $e$ is black}\}|\ge \sum_{e\in E_v}X_e.
\end{equation}
Next, we are interested in bounds on $\Prob{}{X_e = 1}$. Assume $e$ is incident in $u$ and let $S$ be the set of edges incident in $u$ observed up to time $t$. 
Then, from \Cref{fa:perm}, the sequence in which these edges are observed is just a random permutation of $S$. 
This immediately implies that, if $e$ is incident to vertex $u\in L_1(v)$, then  
\[
    \Prob{}{X_e = 1} = \frac{\deg_u - \left\lfloor\frac{\varphi}{1 + \varphi}\deg_u\right\rfloor}{\deg_u}\ge\frac{1}{1 + \varphi}.
\]
Together with \eqref{eq:apx_balls} this yields:
\[
    \Expec{}{|\apxball_2(v)|}\ge 1 + |L_1(v)| + \frac{1}{1 + \varphi}|L_2(v)| \geq \frac{1}{1+\varphi}\vert \ball_2(v) \vert.
\]
We next show that $\sum_{e\in E_v}X_e$ is concentrated around its expectation when $|L_2(v)|$ is large enough, which implies that $|\apxball_2(v)|$ is concentrated around a value close to $|\ball_2(v)|$ in this case. The main technical hurdle here is that the $X_e$'s are correlated (albeit mildly, as we shall see). To prove concentration, we resort to Martingale properties of random edge sequences to apply the method of (Average) Bounded Differences \cite{dubhashi2009concentration}. In order to do this, we need bounds on $\Prob{}{X_e = 1\vert X_f = 1}$ and $\Prob{}{X_e = 1\vert X_f = 0}$, for $e, f\in E_v$. Assume again that $e$ is incident in $u\in L_1(v)$ and that $S$ is the set of edges incident in $u$ observed up to time $t$. Assume first that $f$ is also incident in $u$ and that, without loss of generality, $f$ is the $i$-th edge to appear among those in $S$. $X_f = 1$ implies $i\le\deg_u - \left\lfloor\frac{\varphi}{1 + \varphi}\deg_u\right\rfloor$. On the other hand, for any such choice for $f$'s position in the sequence, Fact \ref{fa:perm} implies that $e$ will appear in a position $j$ that is sampled uniformly at random from the remaining ones, so that $\Prob{}{X_e = 1\vert X_f = 1} = \frac{\deg_u - \left\lfloor\frac{\varphi}{1 + \varphi}\deg_u\right\rfloor - 1}{\deg_u - 1}$ in this case. With a similar argument, it can be seen that $\Prob{}{X_e = 1\vert X_f = 0} = \frac{\deg_u - \left\lfloor\frac{\varphi}{1 + \varphi}\deg_u\right\rfloor}{\deg_u - 1}$. Intuitively and unsurprisingly, the events $(X_e = 1)$ and $(X_f = 1)$ are negatively correlated, while $(X_e = 1)$ and $(X_f = 0)$ are positively correlated. This allows us to conclude that $\Prob{}{X_e = 1\vert X_f = 1}\le \Prob{}{X_e = 1\vert X_f = 0}$ and 

\[
    \Prob{}{X_e = 1\vert X_f = 0} - \Prob{}{X_e = 1\vert X_f = 1}\le\frac{1}{\deg_u - 1}.
\]
Assume next that $f$ is not incident in $u$. Again from Fact \ref{fa:perm}, in this case $f$ has no bearing on the relative order in which edges incident in $u$ appear, so that $\Prob{}{X_e = 1\vert X_f = 0} = \Prob{}{X_e = 1\vert X_f = 1} = \Prob{}{X_e = 1}$. Now, without loss of generality, suppose that $f = (z, w)$, with $z\in L_1(v)$, so that $w\in C_{z}$. Denote by $E_v(z)$ the subset of edges in $E_v$ with one end point in $C_{z}$. Moving to conditional expectations we have
\begin{align*}
    &\Expec{}{\sum_{e\in E_v}X_e \, \vert\,  X_f = 0} - \Expec{}{\sum_{e\in E_v}X_e \, \vert\, X_f = 1} \\
    &= \sum_{e\in E_v}\left(\Prob{}{X_e = 1 \, \vert\, X_f = 0} - \Prob{}{X_e = 1 \, \vert\, X_f = 1}\right)\\
    &= \sum_{e\in E_v \setminus E_v(z)}\left(\Prob{}{X_e = 1 \, \vert\, X_f = 0} - \Prob{}{X_e = 1 \, \vert\, X_f = 1}\right) \\
    &+ \sum_{e\in E_v(z)}\left(\Prob{}{X_e = 1 \, \vert\, X_f = 0} - \Prob{}{X_e = 1 \, \vert \, X_f = 1}\right)\\
    &\le\frac{|C_{z}|}{\deg_z - 1}\le 1,
\end{align*}
where the third inequality follows from the definition of $C_z$, since $f$ is incident in $z$, while the last inequality follows since $|C_z|\le \deg_z - 1$ for every $z\in L_1(v)$, because one of the edges incident in $z$ is by definition the one shared with $v$.

We can therefore apply \cite[Definition 5.5 and Corollary 5.1]{dubhashi2009concentration}, with $c\le |L_2(v)|$ to obtain, for every $\alpha > 0$:
\begin{align*}
    &\Prob{}{\Expec{}{\sum_{e\in E_v}X_e} - \sum_{e\in E_v}X_e > \alpha\Expec{}{\sum_{e\in E_v}X_e}}\le e^{-\frac{2\alpha^2|L_2(v)|}{(1 + \varphi)^2}},
\end{align*}
where in the right hand side we also used the bound $\Expec{}{\sum_{e\in E_v}X_e}\ge\frac{1}{1 + \varphi}|L_2(v)|$ we showed earlier.
Finally, we recall \eqref{eq:apxball_det} and \eqref{eq:apx_balls} to conclude that $|\apxball_2(v)|\ge \frac{1 - \alpha}{1 + \varphi}|L_2(v)|$ with (at least) the same probability.
\end{proof}

\Cref{thm:random_seq_quality} easily implies approximation bounds on indices that depend on the union and/or intersection of $2$-balls. For example, we immediately have the following approximation bound on the Jaccard similarity between any pair of $2$-balls.

\begin{corollary}\label{cor:jacc}
  Under the same assumptions as \Cref{thm:random_seq_quality}, at any time step $t \geq 1$ and for any pair of vertices $u,v \in V$, \lazyscheme$(\varphi,k)$ guarantees the following approximation of the Jaccard similarity between $\ball_2(u)$ and $\ball_2(v)$ with probability at least $1 - e^{-\frac{2\alpha^2|L_2(u)|}{(1 + \varphi)^2}} - e^{-\frac{2\alpha^2|L_2(v)|}{(1 + \varphi)^2}}$: 
    \begin{equation}\label{eq:jacc_apx}
        \textstyle \dfrac{\jacc(\ball_2(u),\ball_2(v))}{1-2\varepsilon'} \geq \jacc(\apxball_2(u), \apxball_2(v)) \geq (1-\varepsilon')\jacc(\ball_2(u),\ball_2(v)) - \varepsilon',
    \end{equation}
    where $\varepsilon' = \frac{\varphi + \alpha}{1 + \varphi}$.
\end{corollary}
\begin{proof}
It is easy to see that $|\apxball_2(u)|\ge (1 - \varepsilon')|\ball_2(u)|$ and $|\apxball_2(v)|\ge (1 - \varepsilon')|\ball_2(v)|$ together imply \eqref{eq:jacc_apx} deterministically. The result then immediately follows from \Cref{thm:random_seq_quality} and a union bound on the events $(|\apxball_2(u)| < (1 - \varepsilon')|L_2(u)|)$ and $(|\apxball_2(v)| < (1 - \varepsilon')|L_2(v)|)$.
\end{proof}

\section{Adversarial edge  sequences}\label{sec:gammaok}
We next study our lazy-update algorithm in an adversarial framework. We show in \Cref{ssec:lowerbound} that if the adversary can \textit{both}: i) choose a worst-case  graph $G$ \textit{and} ii)   submit $G$ according to an \textit{adaptive} sequence of edge insertions, then it is possible to prove a strong lower bound on the achievable update-time/approximation trade-off of the whole parameterized scheme $\lazyscheme (\varphi,k)$.

On the other hand, in \Cref{ssec:gammaok} we provide a \textit{necessary} condition for the adversarial, worst-case framework above: the \textit{girth} \cite{diestel2024graph} of $G$ must be smaller than 5. More precisely, $G$ must contain an unbounded number of triangles or cycles of length 4. We do this by showing that for a suitable parameter setting, algorithm $\lazyscheme (\varphi,k)$ achieves almost-optimal trade-offs even on adversarial edge insertion sequences, for every graph that has a bounded number of such small cycles (see \Cref{def:gammaok} for a formal definition of this class of graphs).

\subsection{A lower bound for adversarial sequences} \label{ssec:lowerbound}
The lower bound for the adversarial framework described above is formalized in the following result on the approximation ratio (see Def. \ref{def:coverage})

\begin{theorem}\label{thm:lower}
    For every $\varphi \in [0,1]$, and integer $k \geq 0$, if  \lazyscheme$(\varphi,k)$ has approximation ratio $\rho \ge 1$, then it must have an amortized update cost  of $\Omega(\Delta/\rho^3)$, where $\Delta$ is the maximum degree of the graph.
\end{theorem}\label{le:lb1}

\begin{figure}[ht]
    \centering
    \includegraphics[width=0.66\linewidth]{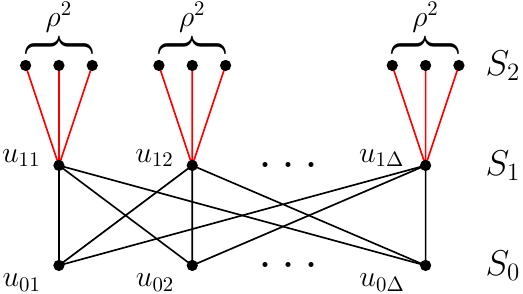}
    \caption{Black edges are present at $t = 0$, while red ones are inserted in the interval $\{1, 2,\ldots , \Delta \rho^2\}$. At time $t > 0$, an edge with one endpoint in $u_{1t\mod\Delta}$ and the other in a distinct 0-degree vertex in $S_2$ is added.}
    \label{fig:lb1}
\end{figure}

\begin{proof}
Fix $\rho \ge 1$, and assume that \lazyscheme$(\varphi,k)$ has an approximation ratio of at most $\rho$. We will show that there exist an initial graph $G^{(0)}$ with degree $\Delta$ and a sequence of edge insertions against which \lazyscheme$(\varphi,k)$ must incur an amortized update time of $\Omega(\Delta/\rho^3)$.

Note that for $k>0$, the algorithm is randomized. In order to address this, we prove our lower bound for every possible realization of the randomness used by the algorithm. Therefore, we assume the values of the random bits used by \lazyscheme$(\varphi,k)$ are fixed arbitrarily (and optimally) and we assume henceforth that the behavior of the algorithm is completely deterministic. 

The initial graph $G^{(0)}$ consists on $2 \Delta$ vertices forming a complete bipartite graph with sides $S_0=\{u_{01},\dots,u_{0\Delta}\}$ and $S_1=\{u_{11},\dots,u_{1\Delta}\}$, along with an additional set $S_2$ of $\Delta \rho^2$ isolated vertices (see \Cref{fig:lb1}).
The sequence of edge insertions is defined as follows: 
for each vertex in $S_1$, we insert $\rho^2$ new edges. Each of these $\Delta\rho^2$ edges connects a vertex in $S_1$ to a previously isolated vertex in $S_2$.
% We insert $\rho^2$ new edges incident to each vertex in $S_1$. Each of these $\Delta \rho^2$ new edges has an endpoint in $S_1$, while the other is a previously $0$-degree vertex in $S_2$. 

%These edges are inserted in a round-robin fashion: we first insert an edge incident to $u_{11}$, then one incident to $u_{12}$ up to one incident to $u_{1\Delta}$, then again another incident to $u_{11}$ and so on. We can view the sequence of insertions as organized in $\rho^2$ rounds of $\Delta$ steps each.
%In the $i$-th step of the $j$-th round one edge is added from $u_{1i}$ to a different vertex belonging to $S_2$.

Consider the time instant right after all edge insertions. 
Since we assumed that the algorithm guarantees an approximation ratio of $\rho$, it holds that for every vertex $u \in S_0$, $|\apxball_2(u)| \ge \frac{1}{\rho} |\ball_2(u)|=\frac{1}{\rho} (2\Delta+\Delta \rho^2) = \Delta \rho + 2\Delta/\rho$. This implies that after all edge insertions, $u$ must be aware of at least $\Delta \rho + 2\Delta/\rho - 2\Delta=\Delta\rho -2 \Delta(1-1/\rho)$ vertices from $S_2$. 

We say that there is a \emph{message} from $v$ to $u$ if vertex $v$ performs a union operation of the form $\apxball_2(u) \gets \apxball_2(u) \cup \apxball_1(v)$.

Since, at any time, every vertex $v \in S_1$ is adjacent to at most $\rho^2$ vertices in $S_2$, it must be that each $u \in S_0$ must have received at least 
$\frac{\Delta\rho -2 \Delta(1-1/\rho)}{\rho^2}=\Omega(\Delta/\rho)$
messages from vertices in $S_1$. As a consequence, the total number of messages are at least $\Omega(\Delta^2/\rho)$. As the number of insertions is $\Delta \rho^2$, the amortized update cost per insertion is at least $\frac{\Omega(\Delta^2/\rho)}{\Delta \rho^2}=\Omega(\Delta/\rho^3)$.
\end{proof}

We have special cases as corollaries. We need amortized update cost $\Omega(\Delta)$ if we want $\rho = O(1)$, $\Omega(\sqrt[4]{\Delta})$ if we want $\rho = O(\sqrt[4]{\Delta})$ and so on. 

\begin{remark}
    The lower bound in \Cref{le:lb1} in fact holds for a wider class of algorithms. Informally speaking, this class includes any \textit{local} algorithm that limits its online updates to the 2-hop neighbors of $u$ and $v$ only. Making this claim more formal requires addressing several technical issues that are outside the scope of the present work. 
\end{remark}

%\subsection{Adversarial edge sequences of large girth}
\subsection{Locally \texorpdfstring{$\gamma$}{gamma}-sparse graphs} \label{ssec:gammaok}

In this section, we provide the characterization of a class of graphs for which our lazy-update approach always guarantees good amortized cost/approximation trade-offs, even under the assumption of adversarial edge insertion sequences.

% In this section, we analyze the performance of our lazy-update approach over a class of graphs that satisfy a property of ``local-sparsity''.
Given a graph $G(V,E)$ and a subset $V' \subseteq V$, we denote by $G[V']$ the subgraph induced by $V'$. 
Informally, a graph is \gammaok\ if every node in $\ball_2(u) \setminus \{u\}$ has roughly at most $\gamma$ neighbors in $\lset_1(u)$. This notion can be formalized as follows. 

%We will prove that, for constant values of $\gamma$, it is possible to obtain a $(1-\varepsilon)$-covering with amortized update  cost $O(\frac{1}{\varepsilon})$.

\begin{definition}[\gammaok\ graphs] \label{def:gammaok}
 Let $\gamma \in 
 \{ 0,1, \ldots , n-2\}$. A graph $G(V,E)$ is said \gammaok\ if for each vertex $u \in V$:
    (i) $\forall v \in \lset_1(u)$ the degree of $v$ in $G[\lset_1(u)]$ is at most $\gamma$, and (ii) $\forall w \in \lset_2(u)$ the degree of $w$ in $G[\lset_1(v) \cup \{ w \}]$ is at most $\gamma+1$.
\end{definition}

%In the following, we make an abuse of notation and say that a dynamic graph $G$ is \gammaok if any $G^{(t)}$ is at most \gammaok, for any $t$.

Observe that the class of \gammaok\ graphs grows monotonically with $\gamma$, including all possible graphs for $\gamma=n-2$, while the most restricted class is obtained for $\gamma=0$. It is interesting to note that \gammaok\ graphs are not necessarily sparse in absolute terms. For example, for $\gamma=0$, the class coincides with the well-known class of graphs with \emph{girth} at least $5$: these graphs can have up to $\Theta(n^{\frac{3}{2}})$ edges assuming Erd{\"o}s' Girth Conjecture \cite{erdos1965some}
%(the proof of such equivalence is given in the full version~\cite{becchetti2025approximate2hopneighborhoodsincremental}).
(the proof of such equivalence is given in \Cref{apx:gamma-ok-deterministic}).

A first, preliminary analysis of our lazy-update approach considers the deterministic version of \Cref{alg:det_thresh}, i.e., when $k = 0$. It turns out that this version achieves an approximation ratio of  $\frac{\gamma + 1}{1-\varepsilon}$ and amortized cost $O(1/\varepsilon)$ 
%(see full version~\cite{becchetti2025approximate2hopneighborhoodsincremental}).
(see \Cref{apx:gamma-ok-deterministic}).
So, the approximation accuracy decreases linearly in the parameter $\gamma$. 
Interestingly enough, we instead show that a suitable number of random light updates allows \Cref{alg:det_thresh} to perform much better than its deterministic version. This is the main result of this section and it is formalized in the following theorem.

\begin{theorem}\label{thm:gamma-ok-main}
Let $\varepsilon \in (0,1)$, and let $G^{(0)}$ be an initial graph. Consider any sequence of edge insertions that yields a final graph $G$. If $G$ is \gammaok\, \lazyscheme$\left(\varphi =1,nk=\frac{4(\gamma+1)}{\varepsilon}\right)$ has approximation ratio of $\frac{1}{1-\varepsilon}$ and amortized cost $O\left(\frac{\gamma+1}{\varepsilon}\right)$.     
\end{theorem}

%We in fact prove that, by setting $k = \frac{4(\gamma + 1)}{\varepsilon}$ and $\varphi = 1$, \lazyscheme\ achieves an approximation ratio of $\frac{1}{1-\varepsilon}$ and amortized update cost of $O(\frac{\gamma+1}{\varepsilon})$ for \gammaok graphs. 

\subsubsection*{Proof of \Cref{thm:gamma-ok-main}}
% For the remainder of this proof, we define the notion of \emph{quasi-black} edge. Informally, a red edge $(v,w)$, with $v \in \lset_1(u)$ and $w \in \lset_2(u)$, is \textit{quasi-black} for $u$ if $u$ is selected in Line 14 of \Cref{alg:det_thresh} for the subsequent insertion of an edge $(v,w')$, ensuring that $w \in \apxball_2(u)$. More formally: \rem{forse la def formale si può togliere}

% \begin{definition}
% Let $u \in V$, and $v \in \lset_1(u)$. For $i=1,\dots,\rd_v$, let $e_i$ be the $i$-th red edge w.r.t. $v$ inserted in the graph. We say that $e_i$ is a \emph{quasi-black edge for $u$} if $u$ has been randomly selected at least once during the insertions of $e_i,\dots,e_{\rd_v}$ (\Cref{alg:det_thresh} lines 14-16). 
% \end{definition}

For the remainder of this proof, we define the notion of \emph{quasi-black} edge. A red edge $(v,w)$, with $v \in \lset_1(u)$ and $w \in \lset_2(u)$, is said to be \textit{quasi-black} for $u$ if $u$ has been randomly selected by $v$ at \Cref{line:random_selection} of \Cref{alg:det_thresh} \emph{at least once} during or after the insertion of $(v,w)$, ensuring that $w \in \apxball_2(u)$.

In this section, we use $\lrdr_v$ and $\lrd_v$ to denote the number of \emph{quasi-black} and \emph{red} edges, respectively, that connect $v$ to vertices in $\lset_2(u)$. Similarly, we use $\lbdd_v$ to represent the number of \emph{black} edges of $v$ having the other endpoint in $\lset_2(u)$. Notice that $\lrdr_v$ is a random variable that counts how many vertices out of $\lrd_v$ are in $\apxball_2(u)$. We now proceed by first stating a property, whose proof can be found in \Cref{apx:proof_gamma_ok_expect_lowerbound}.

\begin{lemma}\label{le:gamma_ok_expect_lowerbound}
     For each $v \in \lset_1(u)$, we have $\Expec{}{\lrdr_v} \geq \lrd_v - \frac{2(\lbdd_v + \gamma + 1)}{k}$.
\end{lemma}

Now, let $\lbddt$ denote the number of vertices in $\lset_2(u)$ that have at least one black edge from $\lset_1(u)$. Consequently, these vertices are included in $\apxball_2(u)$. We have the following:

\begin{lemma}\label{lemma:gamma_ok_properties}
Let $G=(V,E)$ be \gammaok, and $u \in V$. Then
\begin{align*}
     \lbddt\geq \sum_{v \in \lset_1(u)} \frac{\lbdd_v}{\gamma + 1}.
\end{align*}
\end{lemma}
\begin{proof}
    The inequality  follows from the fact that every node in $\lset_2(u)$ can have at most $\gamma + 1$ neighbors in $\lset_1(u)$.  
\end{proof}
  
We are now ready to prove \Cref{thm:gamma-ok-main}. 

%%% BEGIN OF THE PROOF OF THE THEOREM %%%%%
%\begin{proof}

The amortized update cost follows directly from \Cref{lm:amortized_det_alg}.

For the approximation quality, let us consider any vertex $u \in V$. For technical convenience, we will define a subgraph $\widetilde{G}$ of $G$ by removing suitable edges from $G$, and we establish the following two properties: (i) if $k\ge \frac{2(\gamma+1)}{\varepsilon}$, the \lazyscheme\ guarantees a $(1-\varepsilon)$-covering of $\ball_2(u)$ when the sequence of edge insertion is restricted to edges in $\widetilde{G}$; (ii) property (i) implies that the \lazyscheme\ also guarantees a $(1-\varepsilon)$-covering of $\ball_2(u)$ for $G$, provided that $k \geq \frac{4(\gamma+1)}{\varepsilon}$.

\begin{figure}[h]
    \centering
    \includegraphics[width=.8\linewidth]{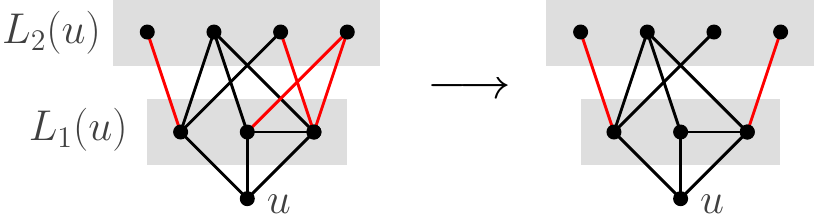}
    \caption{The $2$-hop neighborhood of a vertex $u$ (left), and its corresponding structure in the subgraph $\widetilde{G}$ (right).}
    \label{fig:pruned_graph}
\end{figure}

% We start by defining $\widetilde{G}$ which is obtained from $G$ as follows.
% For each $w \in \lset_2(u)$, if there exists a vertex $v \in \lset_1(u)$ such that edge $(v,w)$ is black for $v$, then we remove all the red edges incident to $w$ that comes from $\lset_1(u)$. Otherwise, we have that all the edges coming from $\lset_1(u)$ are red. In this case we remove all such edges but one. See Figure-?? for an example.\rem{Dobbiamo fare una piccola figura esplicativa.}
The subgraph $\widetilde{G}$ is obtained from $G$ through the following process.
For each vertex $w \in \lset_2(u)$, if there exists a black edge $(v,w)$ with $v \in \lset_1(u)$, then we remove all the red edges incident to $w$ that originate from $\lset_1(u)$.
Otherwise, if all edges from $\lset_1(u)$ to $w$ are red, we retain only one and remove the rest (see \Cref{fig:pruned_graph}).

We now prove property (i). 
We analyze the process at a generic time $t>0$.
We want to prove that $\Expec{}{\vert \apxball_2(u) \vert} \ge (1-\varepsilon)\vert \ball_2(u) \vert$, for any vertex $u \in V$. 
% Since the set $\lset_1(u)$ is always contained in $\apxball_2(u)$, we focus on vertices belonging to $\lset_2(u)$.
Since $\lset_1(u)$ is always included in $\apxball_2(u)$, it is sufficient to prove that $\vert \apxball_2(u) \cap \lset_2(u) \vert \geq (1-\varepsilon) \vert \lset_2(u) \vert$ in expectation.

% Let $\lambda$ and $\hat{\lambda}$ be the number of vertices in $\lset_2(u)$ attached to $\lset_1(u)$ with a red and with a quasi-black edge, respectively.
By construction of $\widetilde{G}$, we have that $\vert \apxball_2(u) \cap \lset_2(u) \vert = \beta + \sum_{v \in \lset_1(v)} \lrdr_v$, while $\vert \lset_2(u) \vert = \beta + \sum_{v \in \lset_1(v)} \lrd_v$.
% Thus, we now want to show that $\lbddt+ \hat{\lambda} \geq (1-\varepsilon)(\lbddt+ \lambda)$ in expectation, i.e.
% \begin{align} \label{eq:errro_bound_2}
%     \hat{\lambda} \ge (1-\varepsilon)\lambda - \varepsilon \lbddt.
% \end{align}
Thus, we want to show that
\begin{align} \label{eq:errro_bound_2}
    \sum_{v \in \lset_1(v)} \lrdr_v \ge (1-\varepsilon)\sum_{v \in \lset_1(v)} \lrd_v - \varepsilon \beta.
\end{align}
%Then, by definition of $\lambda$ and $\hat{\lambda}$ and by \Cref{lemma:gamma_ok_properties}, we get that  \Cref{eq:errro_bound_2} in turn is implied by 
By \Cref{lemma:gamma_ok_properties}, \eqref{eq:errro_bound_2} is true when 
\begin{equation}
\begin{aligned} \label{eq:error_bound_3}
    \sum_{v \in \lset_1(u)}{\lrdr_v} &\ge (1-\varepsilon)\sum_{v \in \lset_1(u)}{\lrd_v} - \frac{\varepsilon}{\gamma + 1}\sum_{v \in \lset_1(u)}{\lbdd_v}\\
    &= \sum_{v \in \lset_1(u)}{\left((1-\varepsilon)\lrd_v - \frac{\varepsilon}{\gamma + 1}\lbdd_v \right)}.
\end{aligned}
\end{equation}
In turn, the inequality in \eqref{eq:error_bound_3} holds in expectation if it holds term-by-term, i.e., when
\begin{equation*}
    \Expec{}{\lrdr_v} \ge (1-\varepsilon)\lrd_v - \frac{\varepsilon}{1+\gamma}\lbdd_v, \;\; \forall v \in \lset_1(u).
\end{equation*}
Clearly, if $\lrd_v = 0$ then $\lrdr_v = 0$ and the inequality holds; thus we focus on the case $\lrd_v > 0$.
From \Cref{le:gamma_ok_expect_lowerbound} we know that $\Expec{}{\lrdr_v} \geq \lrd_v - \frac{2(\lbdd_v + \gamma + 1)}{k}$.
By setting $k \geq \frac{2(\gamma+1)}{\varepsilon}$ we have
\begin{align*}
    \lrd_v - \frac{2(\lbdd_v + \gamma + 1)}{k} \geq \lrd_v - \frac{\varepsilon}{\gamma + 1} \lbdd_v \geq (1-\varepsilon)\lrd_v - \frac{\varepsilon}{\gamma + 1} \lbdd_v,
\end{align*}
therefore proving property (i).

\iffalse
Therefore, we obtain
\begin{align*}
    &\lrd_v - \frac{2(\lbdd_v + \gamma + 1)}{k} \ge (1-\varepsilon)\lrd_v - \frac{\varepsilon}{1+\gamma}\lbdd_v\\
    &\iff \frac{2(\lbdd_v + \gamma + 1)}{k} \leq \varepsilon \lrd_v - \frac{\varepsilon}{1 + \gamma}\lbdd_v \\
    &\iff \frac{k}{2(\lbdd_v + \gamma + 1)} \geq \frac{1+\gamma}{(1+\gamma)\varepsilon \lrd_v + \varepsilon \lbdd_v} \\
    &\iff k \geq \frac{2(1+\gamma + \lbdd_v)(1 + \gamma)}{(1 + \gamma)\varepsilon \lrd_v +\varepsilon \lbdd_v}
\end{align*}
Lastly, observe that
\begin{align*}
    k \ge \frac{2(1+\gamma)}{\varepsilon} \implies k \ge \frac{2(1+\gamma + \lbdd_v)(1 + \gamma)}{(1 + \gamma)\varepsilon \lrd_v + \varepsilon \lbdd_v} =  \frac{2(1+\gamma)}{\varepsilon} \frac{1 + \gamma + \lbdd_v}{(1+\gamma)\lrd_v + \lbdd_v}.    
\end{align*}
Therefore, property (i) is proven.
\fi

% Finally, we prove (ii). Notice that we build $\widetilde{G}$ by removing \emph{only} red edges from $G$.
% On the one hand, this reduces (AUMENTA) the chance to select $u$ at Line 14 in \Cref{alg:det_thresh} as random neighbor of some vertex in $\lset_1(u)$. On the other hand, we reduce the degree of some vertex in $\lset_1(u)$. However, since $\varphi=1$, we have that the red degree of any vertex is at most its black degree. So, in $\widetilde{G}$ the degree of a vertex in $\lset_1(u)$ can at most be halved. In order to manage this we simply double the value of $k$, and the claim follows.
%\end{proof}

Finally, we prove (ii). Notice that we build $\widetilde{G}$ by removing \emph{only} red edges from $G$.
Since $\varphi = 1$, the degrees of the vertices $v \in \lset_1(u)$ in $\widetilde{G}$ are \emph{at most halved} compared to $G$. As a consequence, the probability that $v$ selects $u$ at \Cref{line:random_selection} of \Cref{alg:det_thresh} in $G$ is at most half of the probability we have in $\widetilde{G}$. Therefore, doubling the value of $k$ to $\frac{4(\gamma + 1)}{\varepsilon}$ ensures that the analysis we conducted on $\widetilde{G}$ also holds on $G$, guaranteeing a $(1-\varepsilon)$-covering of $\ball_2(u)$ in expectation on $G$ as well.
\section{Experimental analysis}\label{sec:exp}
The purpose of this experimental analysis, conducted on real datasets, is the validation of our lazy approach along three main axes: i) the qualitative consistency between the theoretical findings from \Cref{sec:detalgo,subse:rand_perm,sec:gammaok} and the actual behavior of our lazy approach on real, incremental datasets;
ii) the accuracy of our approach on two key neighborhood-based queries, namely, size and Jaccard similarity;  iii) its computational savings with respect to non-lazy baselines on medium and large incremental graphs.
An extensive and complete set of results is reported in \Cref{apx:experiments}.

\subsection{Experimental setup}\label{subse:exp_setup}
%\rem{Hardware, measurements etc.}
\paragraph{Platform.} Our experiments were performed on a machine with 2.3 GHz Intel Xeon Gold 5118 CPU with 24 cores, 192 GB of RAM, cache L1 32KB, shared L3 of 16MB and UMA architecture. The whole code is written in \texttt{C++}, compiled with \texttt{GCC 10} and with the following compilation flags: \texttt{-DARCH$\_$X86$\_$64 -Wall -Wextra -g -pg -O3 -lm}.\footnote{Code available at \url{https://github.com/Gnumlab/graph_ball}.}

\paragraph{Algorithms.}
We compared $\lazyscheme(\varphi, k)$ with various combinations $(\varphi, k)$ with  the following baselines: \\
\textit{i)} the exact \textit{algorithm} \ref{algo:naive}, which is not scalable and is only used in the first round of experiments on smaller datasets, in order to isolate the error introduced by lazy updates; \\
\textit{ii)} the naive \textit{sketch-based baseline}, which adopts sketch-based representations of $1$- and $2$-balls, but performs all necessary updates. It corresponds to Algorithm \ref{alg:det_thresh} with $\varphi = 0$ and $k = 0$, but where $1$- and/or $2$-ball unions correspond to merging the corresponding sketches.\footnote{Again, the specific sketch (or sketches) used depends on the neighborhood queries one wants to support.}

With the exception of the first set of experiments (i.e. item (i) above), we compared our algorithm to the sketch-based baseline. For some results, we needed to compute the true value of the parameter of interest for $2$-balls (e.g., size) by executing a suitably optimized BFS. For our \lazyscheme, we used combinations of the following values: $\varphi =  0.1, 0.25, 0.5, 0.75, 1$ and $k = 0, 2, 4, 8$. 
%The baseline (\Cref{algo:naive}) is implemented by setting parameters $\varphi = 0$ and 
%$k = 0$ in our algorithm. 

\paragraph{Implementation details.}
To best assess the performance of algorithms, it would be ideal to minimize the overhead deriving from the management of edge insertions in the graph. We observe that this overhead is the same for all algorithms we tested. Hence, we represented graphs using the \textit{compressed sparse row}  format \cite{Eisenstat1982YaleSM} and, since we knew the edge insertion sequence in advance, we pre-allocated the memory needed to accommodate them, so as to minimize overhead.
For experiments that required hash functions, we used \textit{tabulation hashing} \cite{tabulation-hashing}.

% We used the same graph representation for all algorithms analyzed in this section. Since the overhead due to different graph representations in memory is the same for every algorithm, we implemented the graph using Compressed Sparse Row (CSR) format \cite{Eisenstat1982YaleSM} and pre-allocated the space necessary to the insertion of new edges, so as to mitigate the overhead from dynamic graph management. 

%\rem{studiare possibili versioni più dinamiche dei parametri dipendenti dallo stato del %vertice (es., $k=\log(\vert \apxball_2(v) \vert$ oppure $\varphi = \frac{1}{\log (\rd)}$)}

\paragraph{Datasets.}
We considered real incremental graphs of different sizes, both directed and undirected, available from NetworkRepository \cite{networkrepository}. In our time analysis, we also extracted a large incremental dataset from a large static graph, namely \texttt{soc-friendster} \cite{soc-friendster}, available from SNAP \cite{snapnets}. Following previous work on dynamic graphs \cite{chen2022dynamictree, Hanauer22DynamicSurvey}, we generated an incremental graph by adding edges sequentially and in random order, starting from an empty graph. The main features of these datasets are summarized in \Cref{tab:summary_dynamic_dataset}.

\iffalse
\begin{table*}[h]
    \centering
    \begin{tabular}{l|ccccc|c}
        \toprule
        dataset & $\vert V \vert$ & $\vert E \vert$ & \# insertions & batches & directed & dynamic \\
        \midrule
        comm-linux-kernel-reply & 27,927 & 242,976 & 1,030,000 & 839,643 & \ding{52} & \ding{52} \\
        fb-wosn-friends \cite{fb-wosn-friends} & 63,731 & 817,090 & 1,270,000 & 736,675 & \ding{56} & \ding{52} \\
        ia-enron-email-all & 87,273 & 321,918 & 1,130,000 & 214,908 & \ding{52} & \ding{52} \\
        soc-flickr-growth & 2,302,925 & 33,140,017 & 33,140,017 & 134 & \ding{52} & \ding{52} \\
        soc-youtube-growth & 3,223,589 & 9,376,594 & 12,200,000 & 203 & \ding{52} & \ding{52} \\
        soc-friendster \cite{soc-friendster} & 65,608,366 & 1,806,067,135 & -- & -- & \ding{56} & \ding{56} \\
        \bottomrule

    \end{tabular}
    \caption{Summary table of real networks used in the experiments.}
    \label{tab:summary_dynamic_dataset}
\end{table*}
\fi

\begin{table*}[h]
    \centering
    \begin{tabular}{l|ccc|c}
        \toprule
        Dataset & $\vert V \vert$ & $\vert E \vert$ & Directed & Dynamic \\
        \midrule
        comm-linux-kernel-reply & 27,927 & 242,976 & \ding{52} & \ding{52} \\
        fb-wosn-friends \cite{fb-wosn-friends} & 63,731 & 817,090 & \ding{56} & \ding{52} \\
        ia-enron-email-all & 87,273 & 321,918 & \ding{52} & \ding{52} \\
        soc-flickr-growth & 2,302,925 & 33,140,017 & \ding{52} & \ding{52} \\
        soc-youtube-growth & 3,223,589 & 9,376,594 & \ding{52} & \ding{52} \\
        soc-friendster \cite{soc-friendster} & 65,608,366 & 1,806,067,135 & \ding{56} & \ding{56} \\
        \bottomrule

    \end{tabular}
    \caption{Summary table of real networks used in the experiments.}
    \label{tab:summary_dynamic_dataset}
\end{table*}

\subsection{Results}\label{subse:exp_res}

As we remarked in the introduction, we focused on the case of undirected graphs for ease of exposition %and for the sake of space, 
but our approach extends seamlessly to directed graphs, such as some of the real examples we consider in this section. In such cases, the only caveat to keep in mind is that we define the $h$-ball of a vertex $u$ as the subset of vertices that are reachable from $u$ over a directed path traversing at most $h$ edges.\footnote{One could alternatively define the $h$-ball of $u$ as the subset of vertices from which it is possible to reach $u$ traversing at most $h$ directed edges. We stick to the former definition, which is more frequent in social network analysis.}

\paragraph{Impact of lazy updates.} The goal of our first experiment is twofold: i) assessing the impact of lazy updates on the estimation of $2$-balls; ii) assessing the degree of consistency between the theoretical findings of Sections \ref{subse:rand_perm} and \ref{sec:gammaok} and the actual behavior of our algorithms on real datasets. In order to isolate the specific contribution of lazy updates in the estimation error, we implemented (true and approximate) $1$-balls and $2$-balls losslessly, as dictionaries. This way, the error in $2$-ball size estimation is only determined by our lazy update policy. Since, as we argued elsewhere in the paper, lossless representations of $2$-balls quickly becomes unfeasible for larger datasets, this first experiment was run on $3$ small-medium datasets, namely, \texttt{comm-linux-kernel-reply}, \texttt{fb-wosn-friends} and \texttt{ia-enron-email-all}
%(results for \texttt{comm-linux-kernel-reply} are reported in the full version~\cite{becchetti2025approximate2hopneighborhoodsincremental}).
(results for \texttt{comm-linux-kernel-reply} are reported in \Cref{fig:covering_linux2} in \Cref{apx:experiments}).

\iffalse
In particular, we used four incremental datasets using their respective incremental sequence of insertion. This gives an empirical validation on real update sequences.

We test our solutions w.r.t. their ability to estimate the balls size against real and random incremental sequences. The aim is twofolds: on the one hand, we show that our analysis on random sequences \Cref{sec:rand_worst} works in practice. One the other hand, we show that with real incremental sequences we have the same experimental results, meaning that the analysis we performed is valid for real scenario. We keep track of the error between the true and estimated sizes of two-hop balls over a suitable sample of vertices. For an individual vertex $u$, the error is measured as $\frac{\vert\ball_2(u)\vert}{\vert\apxball_2(u)\vert}$, we then also consider the Root Mean Square Error (RMSE) over the chosen sample of vertices.
\fi

\begin{figure}
    \centering  
    \begin{subfigure}{\linewidth}
        \centering
        \includegraphics[width=.9\linewidth]{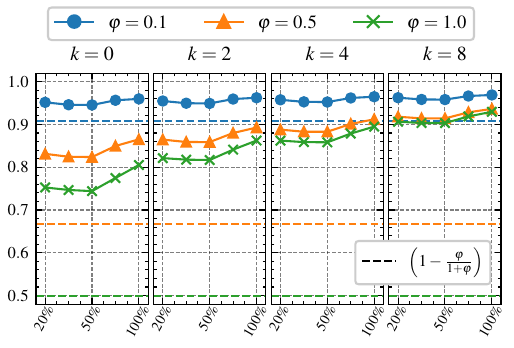}
        \caption{fb-wosn-friends}
        \label{fig:covering_fb}
    \end{subfigure}
    \hfill % Spazio verticale tra le subfigure
    \begin{subfigure}{\linewidth}
        \centering
        \includegraphics[width=.9\linewidth]{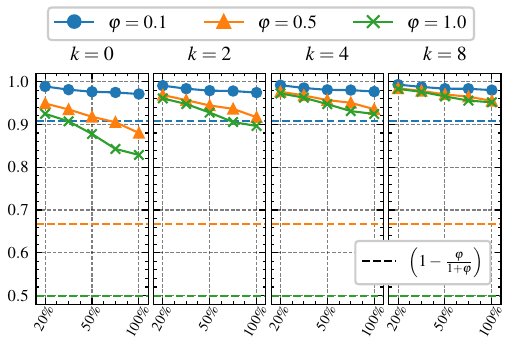}
        \caption{ia-enron-email-all}
        \label{fig:covering_enron}
    \end{subfigure}
    \caption{Average coverage $C_t$ from a network with $20\%$ of its edges to the end of the insertion sequence. Dashed lines show the theoretical expected coverage for random sequences (see \Cref{thm:random_seq_quality}).}
    \label{fig:covering}
\end{figure}

For each of the above graphs, we selected as a sample $5000$ vertices whose $2$-balls are the largest at the end of the edge insertion sequence. For every pair $(\varphi, k)$ of parameter values for \lazyscheme$(\varphi,k)$, we performed $10$ independent runs. Each run is organized into the following steps: 1) the initial graph $G^{(0)}(V,E^{(0)})$ corresponds to the first $20\%$ edge insertions; 2) we measure the coverage (see \Cref{def:coverage}) of each of the $5000$ $2$-balls above by \lazyscheme$(\varphi,k)$ at each of the $100$ equally spaced timestamps, the same in each run. For the generic timestamp $t$, we measure the average coverage
\[
    C_t = \frac{1}{5000}\sum_{i=1}^{5000}\frac{\apxball_i}{\ball_i},
\]
where $\ball_i$ and $\apxball_i$ respectively denote the true and estimated sizes of the $i$-th $2$-ball from the sample. Finally, for each timestamp $t$, we plot the average of the $10$ values of $C_t$ computed in every run. 
The results, summarized in \Cref{fig:covering}, are fully consistent with our theoretical findings from \Cref{subse:rand_perm}. 
At least for the diverse dataset sample considered here, uniform random permutations are a reasonable theoretical proxy of real sequences. More in general, real sequences seem to be rather far from the pathological worst-cases analyzed in \Cref{ssec:lowerbound}, so that the actual behavior of our algorithm is not only in line, but better than our analysis predicts.
We also have an initial insight into the effects of the parameters $\varphi$ and $k$, which will be examined more thoroughly in the subsequent subsections.

\paragraph{Ball size estimation via sketches.} 
The goal of the next round of experiments was assessing the accuracy of our algorithms in $2$-ball size estimation, in the realistic setting in which approximate $1$- and $2$-balls are represented via state-of-art sketches that are based on probabilistic counters \cite{trevisan/646978.711822}. In this case, we have a compound estimation error, arising from both our lazy update policy and the use of probabilistic counters. We ran experiments on the three small-medium sized datasets considered previously, plus two large ones, namely, \texttt{soc-youtube-growth} and \texttt{soc-flickr-growth}. Again, we considered the top-$5000$ largest $2$-balls as sample. The experiments were executed as in the previous case, with the following differences: i) we considered $3$ timestamps, respectively corresponding to $50\%$, $75\%$ and $100\%$ of all edge insertions of each dataset; ii) in this case, $1$- and $2$-balls are not explicitly represented as sets (not even by the baseline). As a result, ball sizes might be overestimated and coverage has no clear meaning.
%\rem{Andy: spiegherei meglio questo cambio di indice statistico: la presenza di sketch %rende possibile anche una stima in eccesso della size....}
We therefore use Mean Absolute Percentage Error (MAPE) to measure accuracy,  defined as follows:
\[
\text{MAPE} = \frac{1}{5000} \sum_{i=1}^{5000} \frac{\vert \ball_i - \apxball_i \vert}{\ball_i},
\]
where $i$ refers to the $2$-ball size of the $i$-th sampled vertex. For each dataset, this index is computed for each of $10$ independent runs at each of the $3$ timestamps we consider.
Results are summarized for all datasets and combinations $(\varphi, k)$ we consider in \Cref{tab:size-quality} for the last timestamp ($100\%$ of edge insertions). Results for other timestamps ($50\%$ and $75\%$) are similar and are omitted for the sake of clarity
%(see full version~\cite{becchetti2025approximate2hopneighborhoodsincremental} for complete results).
(see \Cref{apx:experiments} for complete results).
The main takeaways here are that i) the additional error introduced by the use of sketches (which can be controlled by varying the size of the sketch) is relatively modest; ii) even relatively large values of $\varphi$ and/or small values of $k$ result in performances that are close to those of the baseline that uses sketches to represent balls, but naively performs all light updates. The effect of $k$ is more pronounced when $\varphi$ is large, contributing to a reduction in both error and, to a lesser extent, variance.
Finally, we note that although the effect of increasing $k$ can be achieved by decreasing $\varphi$, our analysis in \Cref{sec:gammaok} suggests that the parameter $k$ provides robustness against worst-case scenarios.
Even though the analysis of the impact of sketches on error is out of the scope of this paper, the results in the previous paragraph and the small difference with the baseline (\Cref{tab:size-quality}) when using the sketches further suggest that the analysis of random sequences may also apply to real sequences.

\begin{table}[ht]
    \centering
    \caption{Mean and standard deviation of absolute percentage errors for $2$-hop neighborhood size estimation. Size estimates were made using the KMV probabilistic counter \cite{trevisan/646978.711822}, with size $32$.
    Queries were made at the end of the insertion sequence.
    }
    \begin{tabular}{lc|ccc|c}
    \toprule
        & $k$ & $\varphi = 0.1$ & $\varphi = 0.5$ & $\varphi = 1$ & baseline\\
    \midrule
    
        \parbox[t]{2mm}{\multirow{4}{*}{\rotatebox[origin=c]{90}{\texttt{linux}}}}

        & $0$ & $0.14 \pm 0.12$ & $0.19 \pm 0.14$ & $0.17 \pm 0.11$ & \multirow{4}{*}{$0.12 \pm 0.10$} \\
        & $2$ & $0.13 \pm 0.11$ & $0.14 \pm 0.10$ & $0.16 \pm 0.11$ & \\
        & $4$ & $0.14 \pm 0.09$ & $0.17 \pm 0.12$ & $0.14 \pm 0.10$ & \\
        & $8$ & $0.14 \pm 0.11$ & $0.14 \pm 0.09$ & $0.13 \pm 0.10$ & \\

        \midrule[.66pt]

        \parbox[t]{2mm}{\multirow{4}{*}{\rotatebox[origin=c]{90}{\texttt{fb-wosn}}}}

        & $0$ & $0.16 \pm 0.12$ & $0.15 \pm 0.11$ & $0.21 \pm 0.12$ & \multirow{4}{*}{$0.14 \pm 0.11$} \\
        & $2$ & $0.13 \pm 0.09$ & $0.15 \pm 0.10$ & $0.17 \pm 0.11$ & \\
        & $4$ & $0.14 \pm 0.11$ & $0.15 \pm 0.10$ & $0.16 \pm 0.11$ & \\
        & $8$ & $0.16 \pm 0.13$ & $0.14 \pm 0.10$ & $0.15 \pm 0.11$ & \\

        \midrule[.66pt]

        \parbox[t]{2mm}{\multirow{4}{*}{\rotatebox[origin=c]{90}{\texttt{enron}}}}

        & $0$ & $0.13 \pm 0.10$ & $0.16 \pm 0.11$ & $0.20 \pm 0.14$ & \multirow{4}{*}{$0.13 \pm 0.11$} \\
        & $2$ & $0.14 \pm 0.11$ & $0.16 \pm 0.12$ & $0.16 \pm 0.12$ & \\
        & $4$ & $0.13 \pm 0.12$ & $0.15 \pm 0.12$ & $0.15 \pm 0.12$ & \\
        & $8$ & $0.13 \pm 0.11$ & $0.14 \pm 0.10$ & $0.16 \pm 0.13$ & \\

        \midrule[.66pt]

        \parbox[t]{2mm}{\multirow{4}{*}{\rotatebox[origin=c]{90}{\texttt{flickr}}}}

        & $0$ & $0.17 \pm 0.14$ & $0.18 \pm 0.12$ & $0.17 \pm 0.11$ & \multirow{4}{*}{$0.17 \pm 0.14$} \\
        & $2$ & $0.16 \pm 0.12$ & $0.12 \pm 0.09$ & $0.14 \pm 0.10$ & \\
        & $4$ & $0.14 \pm 0.09$ & $0.13 \pm 0.10$ & $0.16 \pm 0.10$ & \\
        & $8$ & $0.14 \pm 0.11$ & $0.13 \pm 0.10$ & $0.14 \pm 0.09$ & \\

        \midrule[.66pt]

        \parbox[t]{2mm}{\multirow{4}{*}{\rotatebox[origin=c]{90}{\texttt{youtube}}}}

        & $0$ & $0.15 \pm 0.11$ & $0.15 \pm 0.10$ & $0.24 \pm 0.11$ & \multirow{4}{*}{$0.14 \pm 0.11$} \\
        & $2$ & $0.16 \pm 0.13$ & $0.14 \pm 0.10$ & $0.19 \pm 0.11$ & \\
        & $4$ & $0.13 \pm 0.11$ & $0.13 \pm 0.10$ & $0.15 \pm 0.11$ & \\
        & $8$ & $0.13 \pm 0.10$ & $0.12 \pm 0.09$ & $0.15 \pm 0.11$ & \\
        
    \bottomrule

    \end{tabular}
    \label{tab:size-quality}
\end{table}

\paragraph{Accuracy in Jaccard similarity estimation.} 
With the same goal as in the previous experiment, we now evaluate the quality of our lazy approach policy combined with the use of sketches in another graph mining task: the Jaccard similarity estimation for $2$-hop neighborhoods.
The sketch used to represent the $1$- and $2$- balls is the well-known $h$-minhash \cite{broder1997resemblance}, with $h = 100$ hash functions.

As before, $10$ independent runs were performed for each combination of parameters $\varphi, k$, and the baseline, on each of the previously used datasets.
Errors were measured at $3$ different timestamps, corresponding to $50\%$, $75\%$, and $100\%$ of the edge insertion sequence.

%The $h$-minhash sketch incurs an estimation error of $O(\sqrt{h})$ for the Jaccard similarity \cite{broder2000identifying,broder2001completeness}.
Considering all vertex pairs in the entire graph would be computationally prohibitive, as their number is too large. Moreover, many of these pairs would have an extremely low Jaccard similarity, making it difficult to estimate them with a reasonably small error \cite{broder2000identifying,broder2001completeness}.
Therefore, to better evaluate our algorithm's and baseline's quality, we need vertex pairs with a sufficiently high similarity in their $2$-hop neighborhoods.
%Efficiently sampling such pairs is crucial in this experiment, since analyzing all vertex pairs is computationally infeasible due to their large quantity. In addition, many pairs exhibit very low Jaccard similarity, complicating an accurate estimation.
To address this, we adopted a similar sampling process as before:
we selected the $5000$ vertices with the largest $2$-hop neighborhoods at the end of the edge insertion sequence and randomly chose $1000$ pairs whose similarity is at least $0.2$.
On that sample, we evaluated the MAPE of the Jaccard similarity estimate computed using the $h$-minhash signatures.
\Cref{tab:jacc-similarity-quality} reports the results at the end of the insertion sequence, while results for the other timestamps ($50\%$ and $75\%$) are similar and are omitted for the sake of clarity 
%(see full version~\cite{becchetti2025approximate2hopneighborhoodsincremental} for complete results).
(see \Cref{apx:experiments} for complete results).

\iffalse
Efficiently sampling vertex pairs with a Jaccard similarity of their $2$-hop neighborhoods that is not too small is crucial in this experiment.
Considering all node pairs in the entire graph would be computationally prohibitive, as their number is too large. Moreover, many of these pairs would have an extremely low Jaccard similarity, making it difficult to estimate them with a reasonably small error.
To address this, we adopted the following sampling process:
% Since this experiment requires sampling \emph{pairs} of vertices whose respective $2$-hop neighborhoods have Jaccard similarity that is not too small, the sampling process is as follows:
i) we considered the $5000$ vertices with the largest $2$-hop neighborhoods at the end of the edge insertion sequence, ii) from these, we selected at random the $1\%$ of all $\binom{5000}{2}$ possible pairs, iii) we then filtered this set, retaining only the pairs whose Jaccard similarity of the $2$-balls is $\geq 0.2$, iv) finally, we uniformly sampled $1000$ pairs from the remaining set.
With this approach, we obtained a sample set of $1000$ vertex pairs for each dataset, on which we assess the MAPE of the Jaccard similarity estimation computed using the $h$-minhash signatures.
\Cref{tab:jacc-similarity-quality} reports the results for the first timestamp ($50\%$ \textcolor{red}{(poi diventerà $100\%$ quando finiranno gli esperimenti)} of edge insertions), while results for the other timestamps are similar and are omitted for sake of clarity.
\fi

This experiment further confirms the observations from the previous one: when using sketches to represent the $2$-balls, the errors obtained with our lazy update policy (with appropriate choices of parameters $\varphi,k$) are similar and fully comparable to those of the baseline, which performs all necessary updates.

\begin{table}[h]
    \centering
    \caption{Mean and standard deviation of absolute percentage errors for Jaccard similarity estimation, with $100$ hash functions. Queries were made at the end of the insertion sequence.}
    \begin{tabular}{lc|ccc|c}
    \toprule
        & $k$ & $\varphi = 0.1$ & $\varphi = 0.5$ & $\varphi = 1$ & baseline\\
    \midrule

        \parbox[t]{2mm}{\multirow{4}{*}{\rotatebox[origin=c]{90}{\texttt{linux}}}}

        & $0$ & $0.11 \pm 0.09$ & $0.13 \pm 0.09$ & $0.11 \pm 0.09$ & \multirow{4}{*}{$0.10 \pm 0.08$} \\
        & $2$ & $0.09 \pm 0.07$ & $0.12 \pm 0.09$ & $0.11 \pm 0.09$ & \\
        & $4$ & $0.10 \pm 0.08$ & $0.12 \pm 0.09$ & $0.10 \pm 0.08$ & \\
        & $8$ & $0.09 \pm 0.07$ & $0.12 \pm 0.09$ & $0.10 \pm 0.08$ & \\

        \midrule[.66pt]

        \parbox[t]{2mm}{\multirow{4}{*}{\rotatebox[origin=c]{90}{\texttt{fb-wosn}}}}

        & $0$ & $0.70 \pm 0.25$ & $0.72 \pm 0.24$ & $0.74 \pm 0.23$ & \multirow{4}{*}{$0.70 \pm 0.24$} \\
        & $2$ & $0.70 \pm 0.24$ & $0.70 \pm 0.25$ & $0.72 \pm 0.24$ & \\
        & $4$ & $0.70 \pm 0.24$ & $0.70 \pm 0.24$ & $0.72 \pm 0.24$ & \\
        & $8$ & $0.70 \pm 0.24$ & $0.70 \pm 0.24$ & $0.71 \pm 0.25$ & \\

        \midrule[.66pt]

        \parbox[t]{2mm}{\multirow{4}{*}{\rotatebox[origin=c]{90}{\texttt{enron}}}}

        & $0$ & $0.09 \pm 0.09$ & $0.14 \pm 0.12$ & $0.15 \pm 0.12$ & \multirow{4}{*}{$0.09 \pm 0.09$} \\
        & $2$ & $0.10 \pm 0.09$ & $0.12 \pm 0.10$ & $0.14 \pm 0.12$ & \\
        & $4$ & $0.09 \pm 0.09$ & $0.11 \pm 0.10$ & $0.13 \pm 0.11$ & \\
        & $8$ & $0.10 \pm 0.09$ & $0.10 \pm 0.09$ & $0.11 \pm 0.10$ & \\

        \midrule[.66pt]

        \parbox[t]{2mm}{\multirow{4}{*}{\rotatebox[origin=c]{90}{\texttt{flickr}}}}

        & $0$ & $0.12 \pm 0.09$ & $0.11 \pm 0.09$ & $0.11 \pm 0.09$ & \multirow{4}{*}{$0.11 \pm 0.09$} \\
        & $2$ & $0.11 \pm 0.08$ & $0.10 \pm 0.08$ & $0.11 \pm 0.09$ & \\
        & $4$ & $0.11 \pm 0.08$ & $0.10 \pm 0.08$ & $0.10 \pm 0.08$ & \\
        & $8$ & $0.11 \pm 0.08$ & $0.10 \pm 0.08$ & $0.11 \pm 0.08$ & \\

        \midrule[.66pt]

        \parbox[t]{2mm}{\multirow{4}{*}{\rotatebox[origin=c]{90}{\texttt{youtube}}}}

        & $0$ & $0.12 \pm 0.09$ & $0.11 \pm 0.09$ & $0.17 \pm 0.13$ & \multirow{4}{*}{$0.11 \pm 0.09$} \\
        & $2$ & $0.11 \pm 0.09$ & $0.11 \pm 0.09$ & $0.15 \pm 0.11$ & \\
        & $4$ & $0.11 \pm 0.09$ & $0.13 \pm 0.10$ & $0.16 \pm 0.11$ & \\
        & $8$ & $0.11 \pm 0.09$ & $0.12 \pm 0.09$ & $0.14 \pm 0.10$ & \\

    \bottomrule

    \end{tabular}
    \label{tab:jacc-similarity-quality}
\end{table}

\paragraph{Run time analysis.}
Finally, we measured the running times of our algorithm and the corresponding speed-up with respect to the baseline.
In the following, we report and discuss the results for the task of size estimation, using probabilistic counters.
The performances for the task of Jaccard similarity estimation are analogous, and thus reported in
%the full version~\cite{becchetti2025approximate2hopneighborhoodsincremental}.
\Cref{tab:minhash-time} in \Cref{apx:experiments}.
% The results for the task of Jaccard similarity estimation are analogous, and thus reported in \Cref{tab:minhash-time} in \Cref{apx:experiments}.

\begin{figure}
    \centering
    \includegraphics[width=1\linewidth]{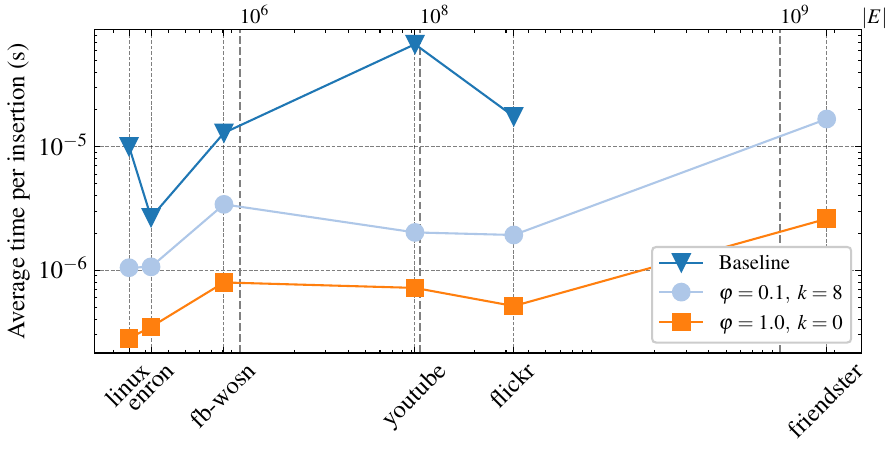}
    \caption{Average time per insertion operation (in seconds) in log-scale.
    The datasets are arranged on the x-axis in increasing order of number of operations, again in log-scale.
    The baseline time for \texttt{soc-friendster} dataset is not reported, since it exceeded a time limit of $36$ hrs.}
    \label{fig:bar_times}
\end{figure}

\Cref{tab:size-time} reports the average speed-up of our algorithm with respect to the naive baseline for the same combinations $(\varphi, k)$ considered previously, for all datasets except \texttt{com-friendster}. Speed-ups are computed in terms of \emph{total update time}, i.e., the total time it takes to process the whole insertion sequence.  \Cref{tab:size-time-friendster} reports the overall processing times on \texttt{com-friendster}, for a subset of the combinations of $\varphi$ and $k$. It should be noted that the baseline did not complete within a reasonable amount of time in this case. Finally, \Cref{fig:bar_times} illustrates the average time cost per operation for the baseline, as well as the slowest and fastest parameter settings of $\varphi$ and $k$ in our algorithm.
These experiments clearly highlight that the number of lazy updates is the crucial factor affecting performance and that our approach is very effective at addressing this problem, resulting in considerable speed-ups that grow with the size of the graph.
These results regarding processing times, in conjunction with previous findings, underscore the significant advantage in terms of speed at the expense of a modest and acceptable reduction in query quality which, due to the necessary use of sketches, is never totally accurate.

\begin{table}[ht]
    \centering
    \caption{Speed up with respect to the baseline, using KMV probabilistic counters \cite{trevisan/646978.711822}, with size $32$.}
    \begin{tabular}{ll|ccccc}
    \toprule
        & \multirow{2}{*}{$k$} & \multicolumn{5}{c}{$\varphi$} \\
        & & $0.1$ & $0.25$ & $0.5$ & $0.75$ & $1$\\
    \midrule

    \parbox[t]{2mm}{\multirow{4}{*}{\rotatebox[origin=c]{90}{\texttt{linux}}}}
    & 0 & 14.39x & 22.73x & 30.27x & 32.36x & 35.65x \\
    & 2 & 12.00x & 18.22x & 22.03x & 21.49x & 23.18x \\
    & 4 & 11.52x & 15.34x & 17.63x & 16.93x & 17.54x \\
    & 8 & 9.57x & 11.86x & 12.80x & 12.03x & 12.22x \\
    \midrule[.66pt]
    
    \parbox[t]{2mm}{\multirow{4}{*}{\rotatebox[origin=c]{90}{\texttt{fb-wosn}}}}
    & 0 & 5.48x & 9.40x & 11.15x & 15.23x & 16.22x \\
    & 2 & 5.22x & 7.35x & 7.16x & 9.21x & 9.67x \\
    & 4 & 4.32x & 6.14x & 6.05x & 7.18x & 7.23x \\
    & 8 & 3.79x & 4.38x & 4.91x & 5.31x & 5.22x \\
    \midrule[.66pt]
    
    \parbox[t]{2mm}{\multirow{4}{*}{\rotatebox[origin=c]{90}{\texttt{enron}}}}
    & 0 & 4.17x & 5.38x & 6.64x & 7.12x & 7.70x \\
    & 2 & 3.61x & 4.11x & 4.68x & 4.82x & 5.27x \\
    & 4 & 3.05x & 3.42x & 3.78x & 3.96x & 4.21x \\
    & 8 & 2.51x & 2.73x & 2.85x & 3.01x & 3.12x \\
    \midrule[.66pt]
    
    \parbox[t]{2mm}{\multirow{4}{*}{\rotatebox[origin=c]{90}{\texttt{flickr}}}}
    & 0 & 13.52x & 19.97x & 26.54x & 31.51x & 34.20x \\
    & 2 & 12.16x & 16.41x & 19.52x & 21.93x & 22.61x \\
    & 4 & 10.76x & 13.96x & 16.35x & 17.17x & 17.79x \\
    & 8 & 9.10x & 11.19x & 12.36x & 12.99x & 13.28x \\
    \midrule[.66pt]
        
    \parbox[t]{2mm}{\multirow{4}{*}{\rotatebox[origin=c]{90}{\texttt{youtube}}}}
    & 0 & 41.61x & 61.80x & 77.08x & 86.50x & 93.73x \\
    & 2 & 38.16x & 51.64x & 60.39x & 64.93x & 65.51x \\
    & 4 & 36.08x & 47.43x & 52.26x & 55.06x & 55.90x \\
    & 8 & 33.33x & 39.60x & 42.73x & 43.61x & 43.82x \\

    \bottomrule

    \end{tabular}
    \label{tab:size-time}
\end{table}

\begin{table}[ht]
    \centering
    \caption{Total running time for ball size estimation using probabilistic counters, for \texttt{com-friendster} dataset. The time for the baseline algorithm is not reported since it exceeded a time limit of $36$ hrs.}
    \begin{tabular}{l|ccccc}
    \toprule
        \multirow{2}{*}{$k$} & \multicolumn{5}{c}{$\varphi$} \\
        & $0.1$ & $0.25$ & $0.5$ & $0.75$ & $1$\\
    \midrule

    $0$ & $5h\; 2'$ & $2h\; 55'$ & $1h\; 53'$ & $1h\; 27'$ & $1h\; 18'$ \\
    $2$ & $6h\; 3'$ & $3h\; 48'$ & $2h\; 59'$ & $2h\; 42'$ & $2h\; 34'$ \\
    $4$ & $6h\; 50'$ & $4h\; 42'$ & $3h\; 56'$ & $3h\; 39'$ & $3h\; 32'$ \\
    $8$ & $8h\; 23'$ & $6h\; 53'$ & $5h\; 54'$ & $5h\; 30'$ & $5h\; 19'$ \\
    
    \bottomrule

    \end{tabular}
    \label{tab:size-time-friendster}
\end{table}

%for a given algorithm, the average update time $T(k)$ after $k$ edge insertions is $
%\frac{\sum_{i=1}^k T_i}{k}$, where $T_i$ denote the time needed by the algorithm to %process the $i$-th edge inserted. In this case, for each dataset, $k$ corresponds to the %total number of edges that are inserted.

\paragraph{A quick reference for parameter setting.}
The trade-off between accuracy, parameter setting and update costs can be summarized as follows: decreasing $\varphi$ increases the frequency of heavy updates (and thus accuracy), while increasing $k$ increases the number of light updates upon a single edge insertion, again improving accuracy.
As our experimental results show, real edge sequences exhibit a behavior that is largely consistent with the random permutation model adopted in the analysis proposed in Section \ref{subse:rand_perm} and far from the adversarial sequences analyzed in Section \ref{sec:gammaok}. In this perspective, the choice of the parameters can be derived by our theoretical results (see \Cref{thm:random_seq_quality} and \Cref{lm:amortized_det_alg}). However, in practice, by running our algorithms on a diverse set of real networks for a multiplicity of parameter settings, we identified two robust settings, corresponding to different choices for the trade-off between accuracy and efficiency. The first setting is $\varphi=0.5$ and $k=0$, which yields a very fast solution with a reasonable accuracy. Instead, when the accuracy requirement is more stringent, we suggest setting $\varphi=0.25$ and $k=2$, which usually provides a negligible error with a reasonable speed-up.

Finally, the setting of the parameters for the sketches we used (in particular, their sizes) was simply driven by i) the trade-off established by the theoretical analyses in \cite{trevisan/646978.711822} and \cite{broder1997resemblance,broder2000identifying,CGPS24} between accuracy and sketch  sizes and ii) the amount of memory we could afford. Ultimately, this resulted in the choices described in Section \ref{subse:exp_setup}, where we used $32$-bit integers.

%We briefly outline principled ways and effective heuristics to choose parameters $\varphi$ and $k$ of our algorithms. We begin with a more principled approach that builds on our theoretical findings from  Section \ref{subse:rand_perm}. We then briefly describe two simple parameters settings, reflecting different choices for the trade-off between accuracy and efficiency.

%As our experimental results show, real edge sequences exhibit a behavior that is largely consistent with the random permutation model adopted in the analysis proposed in Section \ref{subse:rand_perm} and far from the adversarial sequences analyzed in Section \ref{sec:gammaok}. In this perspective, a principled choice of the parameters is mostly driven by our theoretical results.

%The trade-off between accuracy, parameter setting and update costs can be summarized as follows: decreasing $\varphi$ increases the frequency of heavy updates (and thus accuracy), while increasing $k$ increases the number of light updates upon a single edge insertion, again improving accuracy. In particular, the claim of Theorem \ref{thm:random_seq_quality} gives the exact relationship between a desired degree of accuracy described by $\varepsilon$ and the theoretically corresponding choice for $\varphi$. On the other hand, decreasing $\varphi$ and/or increasing $k$ increases the (amortized) update cost per edge insertion. The trade-off described above is quantitatively described in Lemma \ref{lm:amortized_det_alg}.

\section{Node Centrality Ranking}\label{apx:centrality}

In this section, we highlight the effectiveness of our approach in efficiently maintaining a ranking of the vertices of an evolving social network with respect to a popular index of centrality, a key mining task in social network analysis. In this respect, we consider the ranking corresponding to \emph{harmonic centrality} \cite{rochat2009closeness, boldi2014axioms}, a crucial notion in social network analysis, quantifying the importance or influence of a vertex within a network. It is particularly valuable as a refinement of closeness centrality \cite{freeman2002centrality}, addressing some of its limitations \cite{rochat2009closeness,boldi2014axioms}. 

\paragraph{Harmonic centrality.}
Given a graph $G=(V,E)$, the harmonic centrality of a vertex $v \in V$ is defined as
\[
HC(v) = \sum_{u \in V \setminus \{v\}} \frac{1}{d_G(v, u)},
\]
where $d_G(v,u)$ denotes the length (i.e., number of edges) of the shortest path from $v$ to $u$ in $G$, and the value $1/d_G(u,v)$ is conventionally taken to be $0$ when $u$ is not reachable from $v$. It is easy to see that, equivalently, harmonic centrality can be written as follows:
\begin{equation}\label{eq:hc}
    HC(v) = \sum_{i=1}^n\frac{\vert \lset_i(v) \vert}{i},
\end{equation}
where, $\lset_i(v)$ denotes the subset of vertices \emph{exactly} $i$ hops away from $v$. Computing harmonic centrality for all vertices is expensive already in the static setting, in particular, a cost $\Omega(nm)$, corresponding to the naive approach that performs $n$ BFSs, is unavoidable in certain cases \cite{bergamini2019computing}. Of course, problems are magnified in the dynamic setting, where the quest for efficient algorithms, even with linear update times, is an active and promising research area \cite{Hanauer22DynamicSurvey}.

\paragraph{Maintaining harmonic centrality in evolving graphs.}
From an application point of view, the primary interest is often in ranking vertices based on their harmonic centrality or maintaining the top-$k$ highest scoring vertices \cite{bisenius2018computing, crescenzi2020finding, putman2019fast}, rather than calculating the exact values for all of them. In this perspective our lazy approach affords accurate approximation of the actual ranking of the vertices under incremental updates and with update costs that are typically $O(1)$ on real datasets. 

Our approach is straightforward: i) we approximate $HC(v)$ by truncating the summation in \eqref{eq:hc} to its first two terms; ii) rather than maintaining the sets $\lset_1(v)$ and $\lset_2(v)$, we use probabilistic counters to estimate their cardinalities, a strategy that was first explored by Boldi et al. \cite{boldi2013core} in the \emph{static} setting.
More formally, \emph{truncated harmonic centrality} is defined as 
\begin{align*}
HC_2(v)
&= \sum_{u \in \ball_2(v) \setminus\{v\}} \frac{1}{d(v, u)} = \vert \lset_1(v) \vert + \frac{\vert \lset_2(v) \vert}{2}\\
&= \vert \ball_1(v) \vert -1 + \frac{\vert \ball_2(v) \vert - \vert \ball_1(v) \vert}{2}
\end{align*}
Computation of $HC_2(v)$ requires knowledge of the $1$-hop and $2$-hop neighborhoods of $v$ ($\ball_1(v)$ and $\ball_2(v)$ respectively).
To this purpose, it is natural to \emph{approximate} $HC_2(v)$ using the approximate sizes of $\vert \ball_1(v) \vert$ and $\vert \ball_2(v) \vert$ ($\vert \apxball_1(v) \vert$ and $\vert \apxball_2(v) \vert$, respectively), as maintained by our algorithm through probabilistic counters. 
Therefore, we define the \emph{approximate truncated} harmonic centrality $\widetilde{HC}_2(v)$, an estimator of $HC_2(v)$, as follows
\begin{equation}\label{eq:apx_hc}
    HC_2(v) \approx \widetilde{HC}_2(v) = \vert \apxball_1(v) \vert - 1 + \frac{\vert \apxball_{2}(v)\vert - \vert \apxball_{1}(v)\vert}{2}.    
\end{equation}

In the remainder of this section, we show that vertex ranking obtained using $HC(v)$ and $HC_2(v)$ are strongly correlated, justifying our approach to a task that would otherwise be hard to solve in practice for medium or large networks.
This shows that our approach enables accurate approximations of important graph properties in dynamic settings, without incurring the overhead of full neighborhood tracking.

\paragraph{Empirical analysis.}
We perform two experiments to evaluate the effectiveness of our approximation. In both experiments, we compare three different rankings of the nodes, denoted as $\rank^*$, $\rank$ and $\widetilde{\rank}$, each one computed, respectively, according to the following criteria: (i) the exact harmonic centrality, (ii)   the 2-truncated harmonic centrality, and (iii) the approximated truncated harmonic centrality that makes use of our lazy approach. We use the same $3$ small-medium datasets (namely \texttt{comm-linux-kernel-reply}, \texttt{fb-wosn-friends} and, \texttt{ia-enron-email-all}) used in the first experiment in \Cref{subse:exp_res}. The complexity of computing the exact harmonic centrality makes the computation on larger graphs unfeasible.

\begin{figure*}[ht!]
    \centering  
    \begin{subfigure}{.33\linewidth}
        \centering
        \includegraphics[width=\linewidth]{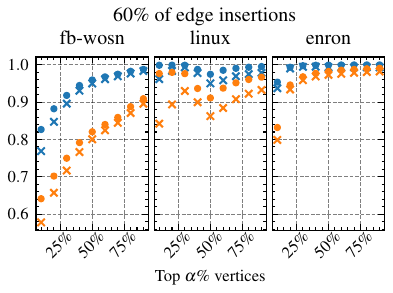}
    \end{subfigure}
    %\hfill % Spazio verticale tra le subfigure
    \begin{subfigure}{.33\linewidth}
        \centering
        \includegraphics[width=\linewidth]{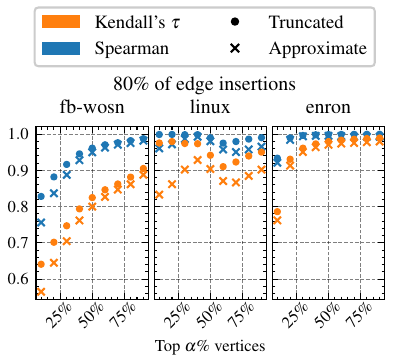}
    \end{subfigure}
    %\hfill % Spazio verticale tra le subfigure
    \begin{subfigure}{.33\linewidth}
        \centering
        \includegraphics[width=\linewidth]{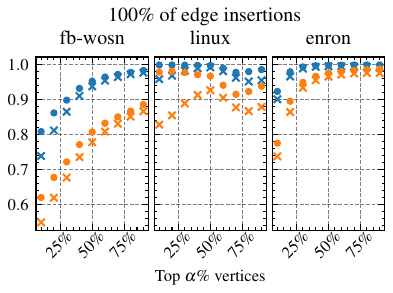}
    \end{subfigure}
    \caption{Spearman's and Kendall's $\tau$ correlation coefficients of the $2$-truncated harmonic centrality $HC_2$ and its approximation $\widetilde{HC}_2$, with respect to the exact harmonic centrality.
    Each point represents the correlation of the top $\alpha\%$ vertices with respect to the ordering induced by $\pi^*$.}
    \label{fig:correlation}
\end{figure*}

In the first experiment (\Cref{fig:correlation}), we focus on the top-ranking nodes of $\rank^*$, for different datasets, in different (time) snapshots (w.r.t. real edge insertion sequences).
For a given value of $\alpha \in \left[ 0,100 \right]$, we select the set $T_{\alpha}^* \subseteq V$ consisting of the top $\alpha\%$ nodes of $\rank^*$.
We then compute the rank correlation between $\rank$ and $\rank^*$, and between $\widetilde{\rank}$ and $\rank^*$, by restricting the rankings to the nodes in $T_{\alpha}^*$. This allows us to assess how well the truncated version and the corresponding approximated one preserve the relative importance of real top central nodes.

The correlations between node rankings is computed using both Spearman's and Kendall's $\tau$  rank correlation coefficients, while as discussed in \Cref{subse:exp_res}, we set the algorithm parameters to $\varphi = 0.25$ and $k = 2$, achieving a good trade-off between update time and approximation quality\footnote{Concerning the sketch, we use the KMV probabilistic counters \cite{trevisan/646978.711822} with size $32$.}. All reported results are averaged over $10$ independent runs, and are summarized in \Cref{fig:correlation}. Overall, the results show that truncated harmonic centrality ($HC_2$) preserves a strong correlation with the exact harmonic centrality ($HC$), confirming its effectiveness as a surrogate metric.
More importantly, our approximation $\widetilde{HC}_2$ achieves correlation scores that are only marginally lower than those of $HC_2$.
This indicates that the approximation introduced by our method is sufficiently accurate to preserve the relative ranking, making it a good and lightweight alternative for tracking centrality in incremental networks.

In the second experiment, we evaluate how well our approach is able to recall/retrieve the top-ranked nodes of $\rank^*$. 
Specifically, for different value of $\alpha \in \{1, 5, 10\}$, we compute the sets $T_\alpha^*,T_\alpha,\widetilde{T}_\alpha$ of the top $\alpha\%$ nodes of $\rank^*$, $\rank$, and $\widetilde{\rank}$, respectively.

We then measure the \emph{recall} of $T_{\alpha}$ and of $\widetilde{T}_\alpha$, defined as the fraction of nodes in $T_\alpha^*$ that also appear in $T_\alpha$ (resp. $\widetilde{T}_\alpha$), \ie $\vert T_\alpha \cap T_\alpha^* \vert / \vert T_\alpha^* \vert$ (resp., $\vert \widetilde{T}_\alpha \cap T_\alpha^* \vert / \vert T_\alpha^* \vert$).
This metric captures how effectively our algorithm identifies the truly central nodes, using only approximate $2$-hop neighborhood information.

\begin{figure*}[ht!]
    \centering
    \begin{subfigure}{0.45\linewidth}
        \centering
        \includegraphics[width=\linewidth]{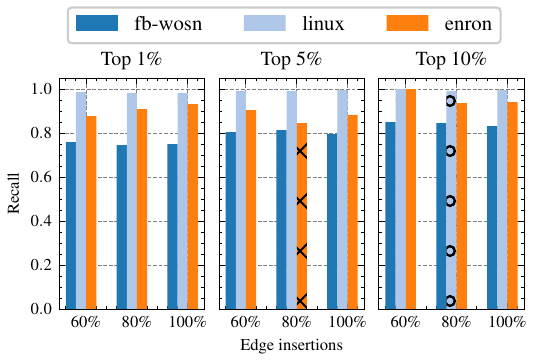}
        \caption{Recall with $HC_2$}
    \end{subfigure}
    %\hfill
    \begin{subfigure}{0.45\linewidth}
        \centering
        \includegraphics[width=\linewidth]{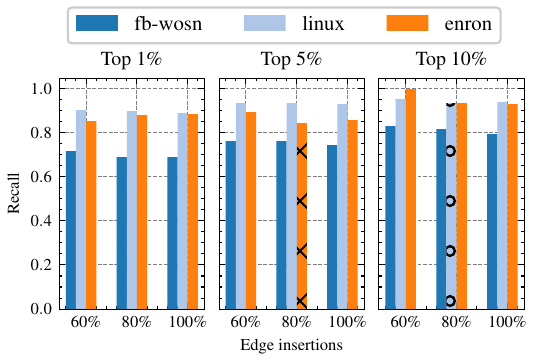}
        \caption{Recall with $\widetilde{HC}_2$}
    \end{subfigure}
    \caption{Recall of the top $1\%$, $5\%$, and $10\%$ vertices using the $2$-truncated harmonic centrality $HC_2$ (on the left), and the approximate harmonic centrality $\widetilde{HC}_2$ (on the right). The x-axis indicate different time snapshot along the edge insertion sequence.}
    \label{fig:recall_top-k}
\end{figure*}

The results are summarized in \Cref{fig:recall_top-k}, and report the average recall over $10$ independent runs.
The reported recall values are consistently high across all datasets and at every point along the edge insertion sequence.
This shows that the top nodes identified by $\widetilde{HC}_2$ closely match those selected by exact harmonic centrality in evolving networks.
Such an accurate recovery of the most central nodes further validates the effectiveness of our algorithm.

\paragraph{Maintaining top-$h$ vertices.}
To conclude this part, we briefly discuss how to efficiently maintain the approximate top-$h$ vertices (e.g., top-$100$ most central vertices) as a continuous query.
In the paragraphs above, we saw how to efficiently maintain (approximate) harmonic centrality values for all vertices of an incremental network using approximate $2$-neighborhood queries. Moreover, our experimental results above show that these approximate values in turn yield a very accurate approximation of the vertices' actual, global ranking. We stress that the involved costs are exactly those analyzed in Section \ref{ssec:detalgo-time-wc}, i.e., a constant amortized cost per edge insertion for typical choices of $\varphi$ and $k$. 

Maintaining the approximate top-$h$ vertices (e.g., top-$100$ most central vertices) as a continuous query can be done efficiently at this point. A standard approach is to use a min-heap to maintain the top-$h$ vertices in terms of their (approximate) harmonic centrality values. As discussed in Section \ref{sec:detalgo}, we manage the insertion of a new edge $(u, v)$ by using the lazy approach which updates the information only of a subset of vertices, say $S$. 
%we at most $\deg_u + \deg_v$ vertices will see their $2$-balls (and their approximate harmonic centrality values) affected (let us denote by $S$ this set of vertices, for simplicity). 
After that, we update the min-heap as follows: In sequence, we consider each vertex $x\in S$. If $x$ belongs to the heap, we update its key to the new value. Otherwise, we compare $x$ against the min-heap's current minimum. If $x$'s updated centrality exceeds the current minimum's then i) the current minimum is removed from the min-heap and ii) $x$ is inserted in the min-heap. The overall, worst-case cost of steps i) and ii) is $O(\log h)$, which implies an (essentially negligible)  additive amortized overhead of $O(\log h)$ per edge insertion.

\medskip
\paragraph{Final remark.} We finally note that alternative (exact or approximate) approaches not considered in this study, e.g., based on on-demand computing or approximating the quantities of interest at query time, are clearly unfeasible for (continuous) queries of the type we consider, each time requiring $n$ local computations, one for each vertex of the graph.  

\section{Discussion and outlook}\label{sec:concl}
We showed that relatively simple, lazy update algorithms can play a key role to efficiently support queries over multi-hop vertex neighborhoods on large, incremental graphs.
At the same time, this work leaves a number of open questions that might deserve further investigation. A first, obvious direction is extending our approach to handle edge deletions.
Actually, a simple strategy to manage a deletion of an edge $(u,v)$ is to recompute the 2-balls of the $O(\deg_u+\deg_v)$ affected vertices via 2-layer BFS visits. The cost of this straightforward approach can be amortized on the entire sequence of edge updates whenever deletions are rare. However, the general case is much more challenging: while our approach is potentially useful, one of the main problems here is handling deletions when compact, sketch-based data structures are used to represent $1$- and $2$-balls. Some recent contributions address similar issues in dynamic data streams \cite{BSS20,CGPS24}, but extending our analyses to this general case does not seem straightforward. Another interesting direction is investigating strategies to handle queries over $h$-balls when $h > 2$, for example maintaining their sizes under dynamic updates. In this case, each edge addition/deletion potentially has cascading effects over $h$-hops. Optimizing (amortized) update costs in this general setting does not seem trivial and we conjecture that a dependence on $h$ might be necessary. Finally, we remark that our algorithms are inherently local, i.e., whether or not to perform updates involving any vertex $v$ only depends on $v$'s immediate neighborhood. As a consequence, a potentially interesting avenue for further research is to investigate distributed variants of our approach, possibly with massive parallel architectures in mind \cite{lee2012parallel}.
%\input{trunk/appendix}
%\input{trunk/rand_worst}
%\input{trunk/notes}
%\nocite{*}
\bibliographystyle{plain}

\clearpage
\newpage
\balance
\bibliography{pool}

\clearpage
\newpage
\appendix

\section{Proofs from Section~\ref{sec:gammaok}} \label{app:gamma}

\subsection{On the girth of locally \texorpdfstring{$\gamma$}{gamma}-sparse graphs}
\begin{figure}[h]
    \centering
    \begin{subfigure}[b]{0.35\linewidth}
            \centering
            \includegraphics[width=\linewidth]{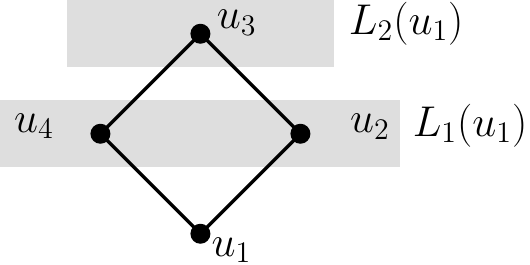}
            \caption{}
            \label{subfig:girth1}
    \end{subfigure}
    \begin{subfigure}[b]{0.6\linewidth}
            \centering
            \includegraphics[width=\linewidth]{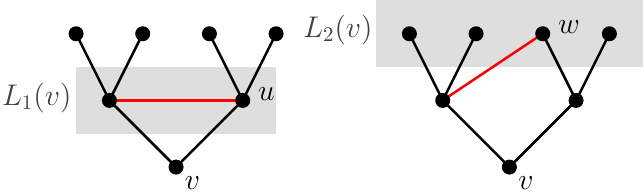}
            \caption{}
            \label{subfig:girth2}
    \end{subfigure}%
    \caption{}
    \label{fig:example_girth}
\end{figure}

\begin{lemma}\label{lemma:girth_rev}
    Let $G = (V,E)$ be an undirected graph with girth $g(G)$.
    Then $G$ is \ok{0} if and only if $g(G) \geq 5$.
\end{lemma}
\begin{proof}
    We first prove that if $G$ is \ok{0} then $g(G)$ must be at least $5$.
    In order to prove that, we simply negate the statement and prove that if $G$ has girth $<5$ then $G$ can not be \ok{0}.
    Without loss of generality, assume that $g(G) = 4$ (the case $g(G) = 3$ is similar).
    Then there must exist a cycle $C = (u_1, u_2, u_3, u_4)$ of $4$ vertices.
    It is simple to see that $u_2,u_4 \in \lset_1(u_1)$ and $u_3 \in \lset_2(u_1)$.
    Since $u_3$ is a neighbor of both $u_2$ and $u_4$, the degree of $u_3$ in the subgraph $G\left[\lset_1(u_1) \cup \{u_4\} \right]$ is at least $2$, hence $G$ is not \ok{0} (see \Cref{subfig:girth1}).
    
    We now prove that if $g(G) \geq 5$ then $G$ must be \ok{0}.
    Again, we negate this statement and prove that if $G$ is not \ok{0} then the girth of $G$ must be less then $5$.
    Let us assume that $G$ is \gammaok, for any $\gamma > 0$, thus it is not \ok{0}.
    Since $G$ is not \ok{0} there exists a vertex $v \in V$ such that at least one of the following properties holds (see \Cref{subfig:girth2}):
    \begin{enumerate}
        \item $\exists u \in \lset_1(v)$ such that the degree of $u$ in $G\left[ \lset_1(v) \right]$ is greater then $0$, or;
        \item $\exists w \in \lset_2(v)$ such that the degree of $w$ in $G\left[ \lset_1(v) \cup \{ w \} \right]$ is greater then $1$.
    \end{enumerate}
    In the first case, we have a cycle of $3$ vertices, then $g(G) = 3$.
    In the second case, we have a cycle of $4$ vertices, then $g(G) = 4$.
    In both cases $g(G) < 5$.
\end{proof}

\subsection{Deterministic lazy-update on \texorpdfstring{$\gamma$}{gamma}-sparse graphs}\label{apx:gamma-ok-deterministic}

\begin{theorem}\label{lemma:gamma-ok-error-bound-balls}
    
Let $\varepsilon \in (0,1)$, and let $G^{(0)}$ be an initial graph. Consider any sequence of edge insertions that yields a final graph $G$. If $G$ is \gammaok, \lazyscheme$(\varphi = \frac{\varepsilon}{1 - \varepsilon},k=0)$ has an approximation ratio of  $\frac{\gamma + 1}{1-\varepsilon}$ and amortized update cost $O(1/\varepsilon)$. 
    
\end{theorem}
\begin{proof}
Recall that $\bd_u$ denotes the black degree of $u$, and that  \Cref{alg:det_thresh} guarantees that $\deg_u$ is at most $(1+\varphi)\bd_u$.
    Then, it is simple to give an upper bound to the size of $\ball_2(u)$, that is $\vert \ball_2(u) \vert \leq 1+ \sum_{v \in \lset_1(u)} (1 + \varphi)\bd_v$.Consider a vertex $v \in \lset_1(u)$. Since $G$ is \gammaok, the number of neighbors of $v$ belonging to $\lset_2(u)$ is at lest $\deg_v - (\gamma+1)$ of which $\bd_v - (\gamma+1)$ must belong to $\apxball_2(u)$. Moreover, a vertex in $\lset_2(u)$ has at most $\gamma+1$ neighbors in $\lset_1(u)$. Therefore: 
    \begin{align*}
    \vert \apxball_2(u) \vert
    &\geq  \bd_u + 1 + \frac{1}{\gamma + 1}\sum_{v \in \lset_1(u)}(\bd_v - (\gamma + 1))\\
    &= \bd_u + 1 + \frac{1}{\gamma + 1}\sum_{v \in \lset_1(u)}\bd_v - \underbrace{\frac{1}{\gamma + 1}\sum_{v \in \lset_1(u)}(\gamma + 1)}_{= \bd_u}\\
    &= 1+ \frac{1}{\gamma + 1}\sum_{v \in \lset_1(u)}\bd_v.
    \end{align*}
  
    As a consequence, $\vert \apxball_2(u) \vert/\vert \ball_2(u) \vert \ge \frac{1}{(1+\varphi)(\gamma+1)}$. By setting $\varphi = \frac{\varepsilon}{1 - \varepsilon}$, and by using \Cref{lm:amortized_det_alg},  the claim follows.
\end{proof}

\subsection{Proof of \Cref{le:gamma_ok_expect_lowerbound}}\label{apx:proof_gamma_ok_expect_lowerbound}
\begin{proof}
Let $e_1, \dots, e_{\ell_v}$ be the \emph{red edges} between $v$ and $\lset_2(u)$, and define the binary random variable $\lrdr_v(i)$ that is equal to $1$ if $e_i$ is a \emph{quasi-black edge} for $u$, $0$ otherwise, for $i = 1, \dots, \lrd_v$. Thus we can express $\lrdr_v = \sum_{i=1}^{\lrd_v} \lrdr_v(i)$, with expectation

\begin{equation}\label{eq:gamma_ok_lb_fact_eq_1}
\begin{aligned}
  \Expec{}{\lrdr_v} & = \sum_{i=1}^{\lrd_v}{\Prob{}{\lrdr_v(i)=1}} = \lrd_v - \sum_{i=1}^{\lrd_v} {\Prob{}{\lrdr_v(i)=0}}.
\end{aligned}
\end{equation}

Without loss of generality, assume that the edges $e_1, \dots, e_{\lrd_v}$ have been inserted at times $t_1 < \dots < t_{\lrd_v}$, respectively.
If $e_i$ is not a quasi-black edge for $u$, then it must be that $u$ is not selected by $v$ at \Cref{line:random_selection} of \Cref{alg:det_thresh}, at times $t_i, t_{i+1},\dots, t_{\lrd_v}$.
This holds with probability 
\begin{equation}\label{eq:gamma_ok_lb_fact_eq_2}
\begin{aligned} 
    &\Prob{}{\lrdr_v(i) = 0}
    \leq \prod_{j=i}^{\lrd_v} \left( 1-\frac{k}{\deg_v^{(t_j)}} \right)
    \leq \prod_{j=i}^{\lrd_v} \left( 1 - \frac{k}{\deg_{v}^{(t_{\lrd_v})}} \right) \\
    &\leq \left( 1-\frac{k}{\lbdd_v + \lrd_v + \gamma + 1}\right)^{\lrd_v - i + 1} 
    \leq \left(1-\frac{k}{2(\lbdd_v + \gamma + 1)}\right)^{\lrd_v - i}.
\end{aligned}
\end{equation}
The third inequality holds since the edges incident to $v$ having endpoints in $L_1(u)$ are at most $\gamma$, while those having endpoints in $L_2(u)$ are exactly $\lbdd_v+ \lrd_v$. Moreover, the last inequality holds because $\lrd_v \leq \rd_v \leq \bd_v \leq \lbdd_v + \gamma + 1$, given the assumption $\varphi = 1$.

By plugging in \eqref{eq:gamma_ok_lb_fact_eq_2} into   \eqref{eq:gamma_ok_lb_fact_eq_1} and we obtain
\begin{align*}
    &\Expec{}{\lrdr_v} \geq \lrd_v - \sum_{i=1}^{\lrd_v}\left( 1-\frac{k}{2(\lbdd_v + \gamma + 1)}\right)^{\lrd_v - i} \\
    &= \lrd_v - \sum_{i=0}^{\lrd_v-1} \left(1-\frac{k}{2(\lbdd_v + \gamma + 1)}\right)^i 
    \leq \lrd_v - \frac{1-\left(1-\frac{k}{2(\lbdd_v+\gamma+1)}\right)^{\lrd_v}}{1-\left(1-\frac{k}{2(\lbdd_v + \gamma + 1)}\right)} \\
    &\geq \lrd_v - \frac{1}{1-\left(1-\frac{k}{2(\lbdd_v + \gamma + 1)}\right)}
    \geq \lrd_v - \frac{2(\lbdd_v + \gamma + 1)}{k}.
\end{align*}
\end{proof}
\newpage
\section{Additional material for Section \ref{sec:exp}}\label{apx:experiments}
In this section, we present the comprehensive set of experimental results, providing an exhaustive overview and offering a detailed picture that completes the material available in \Cref{sec:exp}. %Moreover, we add a concise guide to the setting of the algorithms' parameters.

\begin{figure}[h]
    \centering  
    \begin{subfigure}{\linewidth}
        \centering
        \includegraphics[width=.9\linewidth]{img/plots/covering_fb-wosn-friends_vertical.pdf}
        \caption{fb-wosn-friends}
        \label{fig:covering_fb2}
    \end{subfigure}
    \hfill % Spazio verticale tra le subfigure
    \begin{subfigure}{\linewidth}
        \centering
        \includegraphics[width=.9\linewidth]{img/plots/covering_ia-enron-email-all_vertical.pdf}
        \caption{ia-enron-email-all}
        \label{fig:covering_enron2}
    \end{subfigure}
    \hfill % Spazio verticale tra le subfigure
    \begin{subfigure}{\linewidth}
        \centering
        \includegraphics[width=.9\linewidth]{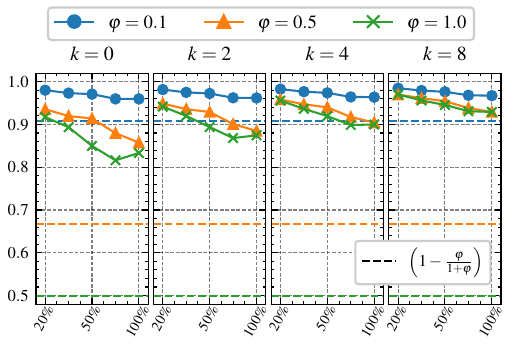}
        \caption{comm-linux-kernel-reply}
        \label{fig:covering_linux2}
    \end{subfigure}
    \caption{Average coverage $C_t$ from a network with $20\%$ of its edges to the end of the insertion sequence. Dashed lines show the theoretical expected coverage for random sequences (see \Cref{thm:random_seq_quality}).}
    \label{fig:covering-full}
\end{figure}

\begin{table}[hb]
    \centering
    \caption{Speed up with respect to the baseline, using minhash sketches, with $100$ hash functions.}
    \begin{tabular}{ll|ccccc}
    \toprule
        & \multirow{2}{*}{$k$} & \multicolumn{5}{c}{$\varphi$} \\
        & & $0.1$ & $0.25$ & $0.5$ & $0.75$ & $1$\\
    \midrule

    \parbox[t]{2mm}{\multirow{4}{*}{\rotatebox[origin=c]{90}{\texttt{linux}}}}
    
    & 0 & 6.73x & 8.01x & 8.67x & 8.96x & 9.08x \\
    & 2 & 6.35x & 7.36x & 7.85x & 8.03x & 8.11x \\
    & 4 & 6.04x & 6.90x & 7.26x & 7.38x & 7.45x \\
    & 8 & 5.52x & 6.13x & 6.35x & 6.40x & 6.45x \\
    
    \midrule[.66pt]
    
    \parbox[t]{2mm}{\multirow{4}{*}{\rotatebox[origin=c]{90}{\texttt{fb-wosn}}}}
    
    & 0 & 2.55x & 3.13x & 3.47x & 3.63x & 3.70x \\
    & 2 & 2.38x & 2.80x & 3.04x & 3.13x & 3.18x \\
    & 4 & 2.27x & 2.60x & 2.76x & 2.81x & 2.86x \\
    & 8 & 2.07x & 2.28x & 2.37x & 2.38x & 2.41x \\
    
    \midrule[.66pt]
    
    \parbox[t]{2mm}{\multirow{4}{*}{\rotatebox[origin=c]{90}{\texttt{enron}}}}
    
    & 0 & 1.85x & 1.99x & 2.07x & 2.10x & 2.12x \\
    & 2 & 1.77x & 1.88x & 1.94x & 1.96x & 1.97x \\
    & 4 & 1.70x & 1.80x & 1.84x & 1.86x & 1.86x \\
    & 8 & 1.59x & 1.65x & 1.67x & 1.68x & 1.68x \\
    
    \midrule[.66pt]
    
    \parbox[t]{2mm}{\multirow{4}{*}{\rotatebox[origin=c]{90}{\texttt{flickr}}}}
    
    & 0 & 12.36x & 17.15x & 20.52x & 22.13x & 23.10x \\
    & 2 & 11.18x & 14.69x & 16.81x & 17.66x & 18.19x \\
    & 4 & 10.36x & 13.11x & 14.63x & 15.22x & 15.59x \\
    & 8 & 9.08x & 10.90x & 11.76x & 12.10x & 12.32x \\
    
    \midrule[.66pt]
    
    \parbox[t]{2mm}{\multirow{4}{*}{\rotatebox[origin=c]{90}{\texttt{youtube}}}}
    
    & 0 & 41.70x & 53.71x & 62.06x & 66.81x & 69.02x \\
    & 2 & 38.74x & 47.56x & 52.71x & 55.24x & 56.28x \\
    & 4 & 36.69x & 43.71x & 47.37x & 49.06x & 49.79x \\
    & 8 & 33.38x & 38.10x & 40.17x & 40.92x & 41.30x \\

    \bottomrule

    \end{tabular}
    \label{tab:minhash-time}
\end{table}

\clearpage
\newpage

\begin{table*}[ht]
    \centering
    \caption{Mean and standard deviation of absolute percentage errors for $2$-hop neighborhood size estimation. Size estimates were made using the KMV counter \cite{trevisan/646978.711822}, with size $32$.
    Each result is done on $10$ independent runs, over a sample of the first $5000$ vertices with the largest $2$-hop neighborhood, after the insertion of $50\%$ of the edges.
    }
    \begin{tabular}{lc|ccccc|c}
    \toprule
        & $k$ & $\varphi = 0.1$ & $\varphi = 0.25$ & $\varphi = 0.5$ & $\varphi = 0.75$ & $\varphi = 1$ & baseline\\
    \midrule

    \parbox[t]{2mm}{\multirow{4}{*}{\rotatebox[origin=c]{90}{\texttt{linux}}}}

    & $0$ & $0.14 \pm 0.12$ & $0.13 \pm 0.09$ & $0.16 \pm 0.11$ & $0.21 \pm 0.13$ & $0.23 \pm 0.13$ & \multirow{4}{*}{$0.12 \pm 0.10$} \\
    & $2$ & $0.13 \pm 0.09$ & $0.15 \pm 0.11$ & $0.14 \pm 0.10$ & $0.17 \pm 0.11$ & $0.20 \pm 0.12$ & \\
    & $4$ & $0.13 \pm 0.09$ & $0.15 \pm 0.11$ & $0.15 \pm 0.10$ & $0.15 \pm 0.11$ & $0.17 \pm 0.11$ & \\
    & $8$ & $0.14 \pm 0.11$ & $0.13 \pm 0.10$ & $0.12 \pm 0.09$ & $0.20 \pm 0.13$ & $0.17 \pm 0.11$ & \\

    \midrule[.66pt]

    \parbox[t]{2mm}{\multirow{4}{*}{\rotatebox[origin=c]{90}{\texttt{fb-wosn}}}}

    & $0$ & $0.15 \pm 0.12$ & $0.16 \pm 0.10$ & $0.19 \pm 0.11$ & $0.21 \pm 0.11$ & $0.27 \pm 0.13$ & \multirow{4}{*}{$0.13 \pm 0.11$} \\
    & $2$ & $0.14 \pm 0.10$ & $0.14 \pm 0.10$ & $0.16 \pm 0.11$ & $0.20 \pm 0.12$ & $0.20 \pm 0.11$ & \\
    & $4$ & $0.14 \pm 0.11$ & $0.14 \pm 0.10$ & $0.15 \pm 0.10$ & $0.17 \pm 0.12$ & $0.17 \pm 0.11$ & \\
    & $8$ & $0.14 \pm 0.12$ & $0.14 \pm 0.11$ & $0.15 \pm 0.11$ & $0.15 \pm 0.11$ & $0.15 \pm 0.10$ & \\

    \midrule[.66pt]

    \parbox[t]{2mm}{\multirow{4}{*}{\rotatebox[origin=c]{90}{\texttt{enron}}}}

    & $0$ & $0.13 \pm 0.10$ & $0.16 \pm 0.11$ & $0.15 \pm 0.10$ & $0.18 \pm 0.12$ & $0.19 \pm 0.13$ & \multirow{4}{*}{$0.14 \pm 0.12$} \\
    & $2$ & $0.14 \pm 0.11$ & $0.14 \pm 0.11$ & $0.16 \pm 0.12$ & $0.15 \pm 0.11$ & $0.16 \pm 0.11$ & \\
    & $4$ & $0.14 \pm 0.12$ & $0.14 \pm 0.11$ & $0.15 \pm 0.10$ & $0.15 \pm 0.12$ & $0.15 \pm 0.12$ & \\
    & $8$ & $0.13 \pm 0.10$ & $0.13 \pm 0.11$ & $0.14 \pm 0.10$ & $0.15 \pm 0.13$ & $0.16 \pm 0.13$ & \\

    \midrule[.66pt]

    \parbox[t]{2mm}{\multirow{4}{*}{\rotatebox[origin=c]{90}{\texttt{flickr}}}}

    & $0$ & $0.19 \pm 0.15$ & $0.16 \pm 0.12$ & $0.18 \pm 0.12$ & $0.16 \pm 0.12$ & $0.17 \pm 0.12$ & \multirow{4}{*}{$0.14 \pm 0.12$} \\
    & $2$ & $0.15 \pm 0.11$ & $0.15 \pm 0.11$ & $0.12 \pm 0.10$ & $0.16 \pm 0.13$ & $0.12 \pm 0.09$ & \\
    & $4$ & $0.13 \pm 0.09$ & $0.15 \pm 0.12$ & $0.14 \pm 0.11$ & $0.15 \pm 0.11$ & $0.15 \pm 0.11$ & \\
    & $8$ & $0.13 \pm 0.11$ & $0.19 \pm 0.15$ & $0.14 \pm 0.12$ & $0.13 \pm 0.11$ & $0.13 \pm 0.09$ & \\

    \midrule[.66pt]

    \parbox[t]{2mm}{\multirow{4}{*}{\rotatebox[origin=c]{90}{\texttt{youtube}}}}

    & $0$ & $0.14 \pm 0.11$ & $0.16 \pm 0.12$ & $0.17 \pm 0.11$ & $0.22 \pm 0.12$ & $0.24 \pm 0.13$ & \multirow{4}{*}{$0.15 \pm 0.12$} \\
    & $2$ & $0.14 \pm 0.10$ & $0.15 \pm 0.11$ & $0.15 \pm 0.11$ & $0.18 \pm 0.12$ & $0.18 \pm 0.11$ & \\
    & $4$ & $0.15 \pm 0.12$ & $0.15 \pm 0.11$ & $0.15 \pm 0.11$ & $0.16 \pm 0.11$ & $0.17 \pm 0.12$ & \\
    & $8$ & $0.15 \pm 0.13$ & $0.14 \pm 0.12$ & $0.14 \pm 0.11$ & $0.16 \pm 0.11$ & $0.16 \pm 0.11$ & \\

    \bottomrule

    \end{tabular}
    \label{tab:size-quality50perc}
\end{table*}

\begin{table*}[ht]
    \centering
    \caption{Mean and standard deviation of absolute percentage errors for $2$-hop neighborhood size estimation. Size estimates were made using the KMV counter \cite{trevisan/646978.711822}, with size $32$.
    Each result is done on $10$ independent runs, over a sample of the first $5000$ vertices with the largest $2$-hop neighborhood, after the insertion of $75\%$ of the edges.
    }
    \begin{tabular}{lc|ccccc|c}
    \toprule
        & $k$ & $\varphi = 0.1$ & $\varphi = 0.25$ & $\varphi = 0.5$ & $\varphi = 0.75$ & $\varphi = 1$ & baseline\\
    \midrule

            \parbox[t]{2mm}{\multirow{4}{*}{\rotatebox[origin=c]{90}{\texttt{linux}}}}

        & $0$ & $0.13 \pm 0.11$ & $0.14 \pm 0.10$ & $0.19 \pm 0.15$ & $0.18 \pm 0.12$ & $0.18 \pm 0.12$ & \multirow{4}{*}{$0.11 \pm 0.09$} \\
        & $2$ & $0.14 \pm 0.11$ & $0.15 \pm 0.10$ & $0.14 \pm 0.10$ & $0.16 \pm 0.11$ & $0.17 \pm 0.11$ & \\
        & $4$ & $0.14 \pm 0.09$ & $0.16 \pm 0.11$ & $0.16 \pm 0.12$ & $0.14 \pm 0.11$ & $0.15 \pm 0.10$ & \\
        & $8$ & $0.15 \pm 0.13$ & $0.14 \pm 0.10$ & $0.13 \pm 0.09$ & $0.19 \pm 0.15$ & $0.14 \pm 0.10$ & \\

        \midrule[.66pt]

        \parbox[t]{2mm}{\multirow{4}{*}{\rotatebox[origin=c]{90}{\texttt{fb-wosn}}}}

        & $0$ & $0.16 \pm 0.12$ & $0.15 \pm 0.10$ & $0.16 \pm 0.11$ & $0.17 \pm 0.11$ & $0.22 \pm 0.12$ & \multirow{4}{*}{$0.14 \pm 0.11$} \\
        & $2$ & $0.13 \pm 0.10$ & $0.14 \pm 0.10$ & $0.15 \pm 0.10$ & $0.17 \pm 0.11$ & $0.18 \pm 0.11$ & \\
        & $4$ & $0.14 \pm 0.11$ & $0.14 \pm 0.11$ & $0.15 \pm 0.10$ & $0.16 \pm 0.13$ & $0.16 \pm 0.11$ & \\
        & $8$ & $0.15 \pm 0.13$ & $0.13 \pm 0.11$ & $0.14 \pm 0.11$ & $0.14 \pm 0.10$ & $0.15 \pm 0.11$ & \\

        \midrule[.66pt]

        \parbox[t]{2mm}{\multirow{4}{*}{\rotatebox[origin=c]{90}{\texttt{enron}}}}

        & $0$ & $0.13 \pm 0.10$ & $0.16 \pm 0.12$ & $0.15 \pm 0.11$ & $0.18 \pm 0.12$ & $0.20 \pm 0.13$ & \multirow{4}{*}{$0.13 \pm 0.11$} \\
        & $2$ & $0.14 \pm 0.11$ & $0.14 \pm 0.10$ & $0.17 \pm 0.12$ & $0.15 \pm 0.11$ & $0.16 \pm 0.12$ & \\
        & $4$ & $0.14 \pm 0.12$ & $0.14 \pm 0.10$ & $0.15 \pm 0.11$ & $0.16 \pm 0.12$ & $0.16 \pm 0.12$ & \\
        & $8$ & $0.13 \pm 0.10$ & $0.15 \pm 0.11$ & $0.14 \pm 0.10$ & $0.14 \pm 0.11$ & $0.16 \pm 0.13$ & \\

        \midrule[.66pt]

        \parbox[t]{2mm}{\multirow{4}{*}{\rotatebox[origin=c]{90}{\texttt{flickr}}}}

        & $0$ & $0.18 \pm 0.15$ & $0.16 \pm 0.11$ & $0.19 \pm 0.12$ & $0.16 \pm 0.11$ & $0.18 \pm 0.11$ & \multirow{4}{*}{$0.17 \pm 0.14$} \\
        & $2$ & $0.14 \pm 0.11$ & $0.16 \pm 0.12$ & $0.12 \pm 0.09$ & $0.16 \pm 0.11$ & $0.14 \pm 0.09$ & \\
        & $4$ & $0.14 \pm 0.09$ & $0.17 \pm 0.13$ & $0.13 \pm 0.10$ & $0.15 \pm 0.11$ & $0.16 \pm 0.10$ & \\
        & $8$ & $0.13 \pm 0.10$ & $0.17 \pm 0.13$ & $0.13 \pm 0.10$ & $0.13 \pm 0.10$ & $0.14 \pm 0.10$ & \\

        \midrule[.66pt]

        \parbox[t]{2mm}{\multirow{4}{*}{\rotatebox[origin=c]{90}{\texttt{youtube}}}}

        & $0$ & $0.14 \pm 0.11$ & $0.15 \pm 0.10$ & $0.19 \pm 0.11$ & $0.21 \pm 0.11$ & $0.24 \pm 0.12$ & \multirow{4}{*}{$0.13 \pm 0.10$} \\
        & $2$ & $0.15 \pm 0.13$ & $0.14 \pm 0.09$ & $0.15 \pm 0.11$ & $0.15 \pm 0.11$ & $0.17 \pm 0.10$ & \\
        & $4$ & $0.14 \pm 0.11$ & $0.15 \pm 0.10$ & $0.16 \pm 0.12$ & $0.14 \pm 0.09$ & $0.16 \pm 0.11$ & \\
        & $8$ & $0.15 \pm 0.12$ & $0.14 \pm 0.11$ & $0.13 \pm 0.09$ & $0.14 \pm 0.10$ & $0.16 \pm 0.10$ & \\

    \bottomrule

    \end{tabular}
    \label{tab:size-quality75perc}
\end{table*}

\begin{table*}[ht]
    \centering
    \caption{Mean and standard deviation of absolute percentage errors for $2$-hop neighborhood size estimation. Size estimates were made using the KMV probabilistic counter \cite{trevisan/646978.711822}, with size $32$.
    Each result is done on $10$ independent runs, over a sample of the first $5000$ vertices with the largest $2$-hop neighborhood, after the insertion of $100\%$ of the edges.
    }
    \begin{tabular}{lc|ccccc|c}
    \toprule
        & $k$ & $\varphi = 0.1$ & $\varphi = 0.25$ & $\varphi = 0.5$ & $\varphi = 0.75$ & $\varphi = 1$ & baseline\\
    \midrule

    \parbox[t]{2mm}{\multirow{4}{*}{\rotatebox[origin=c]{90}{\texttt{linux}}}}

    & $0$ & $0.14 \pm 0.12$ & $0.14 \pm 0.10$ & $0.19 \pm 0.14$ & $0.16 \pm 0.11$ & $0.17 \pm 0.11$ & \multirow{4}{*}{$0.12 \pm 0.10$} \\
    & $2$ & $0.13 \pm 0.11$ & $0.16 \pm 0.10$ & $0.14 \pm 0.10$ & $0.15 \pm 0.11$ & $0.16 \pm 0.11$ & \\
    & $4$ & $0.14 \pm 0.09$ & $0.16 \pm 0.11$ & $0.17 \pm 0.12$ & $0.15 \pm 0.14$ & $0.14 \pm 0.10$ & \\
    & $8$ & $0.14 \pm 0.11$ & $0.13 \pm 0.10$ & $0.14 \pm 0.09$ & $0.21 \pm 0.16$ & $0.13 \pm 0.10$ & \\

    \midrule[.66pt]
    
    \parbox[t]{2mm}{\multirow{4}{*}{\rotatebox[origin=c]{90}{\texttt{fb-wosn}}}}

    & $0$ & $0.16 \pm 0.12$ & $0.15 \pm 0.10$ & $0.15 \pm 0.11$ & $0.17 \pm 0.11$ & $0.21 \pm 0.12$ & \multirow{4}{*}{$0.14 \pm 0.11$} \\
    & $2$ & $0.13 \pm 0.09$ & $0.14 \pm 0.10$ & $0.15 \pm 0.10$ & $0.17 \pm 0.11$ & $0.17 \pm 0.11$ & \\
    & $4$ & $0.14 \pm 0.11$ & $0.14 \pm 0.11$ & $0.15 \pm 0.10$ & $0.16 \pm 0.12$ & $0.16 \pm 0.11$ & \\
    & $8$ & $0.16 \pm 0.13$ & $0.13 \pm 0.11$ & $0.14 \pm 0.10$ & $0.14 \pm 0.10$ & $0.15 \pm 0.11$ & \\

    \midrule[.66pt]

    \parbox[t]{2mm}{\multirow{4}{*}{\rotatebox[origin=c]{90}{\texttt{enron}}}}

    & $0$ & $0.13 \pm 0.10$ & $0.15 \pm 0.12$ & $0.16 \pm 0.11$ & $0.18 \pm 0.12$ & $0.20 \pm 0.14$ & \multirow{4}{*}{$0.13 \pm 0.11$} \\
    & $2$ & $0.14 \pm 0.11$ & $0.14 \pm 0.10$ & $0.16 \pm 0.12$ & $0.15 \pm 0.10$ & $0.16 \pm 0.12$ & \\
    & $4$ & $0.13 \pm 0.12$ & $0.14 \pm 0.10$ & $0.15 \pm 0.12$ & $0.16 \pm 0.12$ & $0.15 \pm 0.12$ & \\
    & $8$ & $0.13 \pm 0.11$ & $0.15 \pm 0.11$ & $0.14 \pm 0.10$ & $0.14 \pm 0.11$ & $0.16 \pm 0.13$ & \\

    \midrule[.66pt]

    \parbox[t]{2mm}{\multirow{4}{*}{\rotatebox[origin=c]{90}{\texttt{flickr}}}}
    
    & $0$ & $0.17 \pm 0.14$ & $0.15 \pm 0.11$ & $0.18 \pm 0.12$ & $0.17 \pm 0.11$ & $0.17 \pm 0.11$ & \multirow{4}{*}{$0.17 \pm 0.14$} \\
    & $2$ & $0.16 \pm 0.12$ & $0.17 \pm 0.12$ & $0.12 \pm 0.09$ & $0.16 \pm 0.11$ & $0.14 \pm 0.10$ & \\
    & $4$ & $0.14 \pm 0.09$ & $0.16 \pm 0.13$ & $0.13 \pm 0.10$ & $0.14 \pm 0.11$ & $0.16 \pm 0.10$ & \\
    & $8$ & $0.14 \pm 0.11$ & $0.16 \pm 0.12$ & $0.13 \pm 0.10$ & $0.13 \pm 0.10$ & $0.14 \pm 0.09$ & \\

    \midrule[.66pt]
    
    \parbox[t]{2mm}{\multirow{4}{*}{\rotatebox[origin=c]{90}{\texttt{youtube}}}}

    & $0$ & $0.15 \pm 0.11$ & $0.13 \pm 0.10$ & $0.15 \pm 0.10$ & $0.23 \pm 0.13$ & $0.24 \pm 0.11$ & \multirow{4}{*}{$0.14 \pm 0.11$} \\
    & $2$ & $0.16 \pm 0.13$ & $0.14 \pm 0.10$ & $0.14 \pm 0.10$ & $0.15 \pm 0.11$ & $0.19 \pm 0.11$ & \\
    & $4$ & $0.13 \pm 0.11$ & $0.16 \pm 0.10$ & $0.13 \pm 0.10$ & $0.15 \pm 0.09$ & $0.15 \pm 0.11$ & \\
    & $8$ & $0.13 \pm 0.10$ & $0.15 \pm 0.11$ & $0.12 \pm 0.09$ & $0.14 \pm 0.10$ & $0.15 \pm 0.11$ & \\
    \bottomrule

    \end{tabular}
    \label{tab:size-quality100perc}
\end{table*}

\begin{table*}[h]
    \centering
    \caption{Mean and standard deviation of absolute percentage errors for Jaccard similarity estimation, with $100$ hash functions. Queries were made after inserting the first $50\%$ of the edges.}
    \begin{tabular}{lc|ccccc|c}
    \toprule
        & $k$ & $\varphi = 0.1$ & $\varphi = 0.25$ & $\varphi = 0.5$ & $\varphi = 0.75$ & $\varphi = 1$ & baseline\\
    \midrule

    \parbox[t]{2mm}{\multirow{4}{*}{\rotatebox[origin=c]{90}{\texttt{linux}}}}

    & $0$ & $0.11 \pm 0.09$ & $0.11 \pm 0.09$ & $0.09 \pm 0.08$ & $0.12 \pm 0.09$ & $0.14 \pm 0.10$ & \multirow{4}{*}{$0.08 \pm 0.07$} \\
    & $2$ & $0.09 \pm 0.07$ & $0.11 \pm 0.09$ & $0.10 \pm 0.08$ & $0.11 \pm 0.08$ & $0.15 \pm 0.11$ & \\
    & $4$ & $0.09 \pm 0.08$ & $0.11 \pm 0.08$ & $0.11 \pm 0.09$ & $0.11 \pm 0.09$ & $0.11 \pm 0.09$ & \\
    & $8$ & $0.09 \pm 0.07$ & $0.11 \pm 0.08$ & $0.10 \pm 0.08$ & $0.11 \pm 0.08$ & $0.11 \pm 0.09$ & \\

    \midrule[.66pt]
    
    \parbox[t]{2mm}{\multirow{4}{*}{\rotatebox[origin=c]{90}{\texttt{fb-wosn}}}}

    & $0$ & $0.80 \pm 0.25$ & $0.80 \pm 0.25$ & $0.80 \pm 0.24$ & $0.81 \pm 0.24$ & $0.81 \pm 0.24$ & \multirow{4}{*}{$0.80 \pm 0.24$} \\
    & $2$ & $0.80 \pm 0.24$ & $0.79 \pm 0.25$ & $0.79 \pm 0.25$ & $0.80 \pm 0.24$ & $0.81 \pm 0.24$ & \\
    & $4$ & $0.79 \pm 0.25$ & $0.79 \pm 0.25$ & $0.80 \pm 0.24$ & $0.80 \pm 0.24$ & $0.80 \pm 0.24$ & \\
    & $8$ & $0.80 \pm 0.24$ & $0.79 \pm 0.25$ & $0.79 \pm 0.25$ & $0.80 \pm 0.24$ & $0.80 \pm 0.24$ & \\

    \midrule[.66pt]
    
    \parbox[t]{2mm}{\multirow{4}{*}{\rotatebox[origin=c]{90}{\texttt{enron}}}}

    & $0$ & $0.09 \pm 0.09$ & $0.11 \pm 0.10$ & $0.12 \pm 0.10$ & $0.13 \pm 0.11$ & $0.14 \pm 0.11$ & \multirow{4}{*}{$0.09 \pm 0.09$} \\
    & $2$ & $0.09 \pm 0.09$ & $0.11 \pm 0.10$ & $0.11 \pm 0.10$ & $0.12 \pm 0.10$ & $0.13 \pm 0.11$ & \\
    & $4$ & $0.09 \pm 0.08$ & $0.10 \pm 0.09$ & $0.11 \pm 0.10$ & $0.12 \pm 0.10$ & $0.12 \pm 0.10$ & \\
    & $8$ & $0.09 \pm 0.09$ & $0.10 \pm 0.09$ & $0.10 \pm 0.09$ & $0.11 \pm 0.09$ & $0.10 \pm 0.09$ & \\

    \midrule[.66pt]

    \parbox[t]{2mm}{\multirow{4}{*}{\rotatebox[origin=c]{90}{\texttt{flickr}}}}

    & $0$ & $0.10 \pm 0.08$ & $0.10 \pm 0.08$ & $0.11 \pm 0.09$ & $0.11 \pm 0.09$ & $0.12 \pm 0.10$ & \multirow{4}{*}{$0.11 \pm 0.09$} \\
    & $2$ & $0.10 \pm 0.08$ & $0.11 \pm 0.09$ & $0.10 \pm 0.08$ & $0.12 \pm 0.09$ & $0.11 \pm 0.09$ & \\
    & $4$ & $0.11 \pm 0.08$ & $0.11 \pm 0.09$ & $0.10 \pm 0.08$ & $0.11 \pm 0.08$ & $0.11 \pm 0.08$ & \\
    & $8$ & $0.11 \pm 0.08$ & $0.11 \pm 0.09$ & $0.10 \pm 0.08$ & $0.12 \pm 0.09$ & $0.11 \pm 0.09$ & \\

    \midrule[.66pt]

    \parbox[t]{2mm}{\multirow{4}{*}{\rotatebox[origin=c]{90}{\texttt{youtube}}}}

    & $0$ & $0.10 \pm 0.09$ & $0.10 \pm 0.09$ & $0.10 \pm 0.08$ & $0.11 \pm 0.09$ & $0.13 \pm 0.11$ & \multirow{4}{*}{$0.12 \pm 0.10$} \\
    & $2$ & $0.11 \pm 0.09$ & $0.11 \pm 0.10$ & $0.11 \pm 0.09$ & $0.16 \pm 0.12$ & $0.12 \pm 0.09$ & \\
    & $4$ & $0.10 \pm 0.08$ & $0.10 \pm 0.08$ & $0.13 \pm 0.11$ & $0.18 \pm 0.13$ & $0.13 \pm 0.09$ & \\
    & $8$ & $0.10 \pm 0.08$ & $0.10 \pm 0.08$ & $0.11 \pm 0.09$ & $0.15 \pm 0.11$ & $0.16 \pm 0.11$ & \\
    
    \bottomrule

    \end{tabular}
    \label{tab:jacc-similarity-quality50perc}
\end{table*}

\begin{table*}[h]
    \centering
    \caption{Mean and standard deviation of absolute percentage errors for Jaccard similarity estimation, with $100$ hash functions. Queries were made after inserting the first $75\%$ of the edges.}
    \begin{tabular}{lc|ccccc|c}
    \toprule
        & $k$ & $\varphi = 0.1$ & $\varphi = 0.25$ & $\varphi = 0.5$ & $\varphi = 0.75$ & $\varphi = 1$ & baseline\\
    \midrule

    \parbox[t]{2mm}{\multirow{4}{*}{\rotatebox[origin=c]{90}{\texttt{linux}}}}

    & $0$ & $0.11 \pm 0.09$ & $0.11 \pm 0.09$ & $0.09 \pm 0.08$ & $0.12 \pm 0.09$ & $0.14 \pm 0.10$ & \multirow{4}{*}{$0.08 \pm 0.07$} \\
    & $2$ & $0.09 \pm 0.07$ & $0.11 \pm 0.09$ & $0.10 \pm 0.08$ & $0.11 \pm 0.08$ & $0.15 \pm 0.11$ & \\
    & $4$ & $0.09 \pm 0.08$ & $0.11 \pm 0.08$ & $0.11 \pm 0.09$ & $0.11 \pm 0.09$ & $0.11 \pm 0.09$ & \\
    & $8$ & $0.09 \pm 0.07$ & $0.11 \pm 0.08$ & $0.10 \pm 0.08$ & $0.11 \pm 0.08$ & $0.11 \pm 0.09$ & \\

    \midrule[.66pt]
    
    \parbox[t]{2mm}{\multirow{4}{*}{\rotatebox[origin=c]{90}{\texttt{fb-wosn}}}}

    & $0$ & $0.80 \pm 0.25$ & $0.80 \pm 0.25$ & $0.80 \pm 0.24$ & $0.81 \pm 0.24$ & $0.81 \pm 0.24$ & \multirow{4}{*}{$0.80 \pm 0.24$} \\
    & $2$ & $0.80 \pm 0.24$ & $0.79 \pm 0.25$ & $0.79 \pm 0.25$ & $0.80 \pm 0.24$ & $0.81 \pm 0.24$ & \\
    & $4$ & $0.79 \pm 0.25$ & $0.79 \pm 0.25$ & $0.80 \pm 0.24$ & $0.80 \pm 0.24$ & $0.80 \pm 0.24$ & \\
    & $8$ & $0.80 \pm 0.24$ & $0.79 \pm 0.25$ & $0.79 \pm 0.25$ & $0.80 \pm 0.24$ & $0.80 \pm 0.24$ & \\

    \midrule[.66pt]
    
    \parbox[t]{2mm}{\multirow{4}{*}{\rotatebox[origin=c]{90}{\texttt{enron}}}}

    & $0$ & $0.09 \pm 0.09$ & $0.11 \pm 0.10$ & $0.12 \pm 0.10$ & $0.13 \pm 0.11$ & $0.14 \pm 0.11$ & \multirow{4}{*}{$0.09 \pm 0.09$} \\
    & $2$ & $0.09 \pm 0.09$ & $0.11 \pm 0.10$ & $0.11 \pm 0.10$ & $0.12 \pm 0.10$ & $0.13 \pm 0.11$ & \\
    & $4$ & $0.09 \pm 0.08$ & $0.10 \pm 0.09$ & $0.11 \pm 0.10$ & $0.12 \pm 0.10$ & $0.12 \pm 0.10$ & \\
    & $8$ & $0.09 \pm 0.09$ & $0.10 \pm 0.09$ & $0.10 \pm 0.09$ & $0.11 \pm 0.09$ & $0.10 \pm 0.09$ & \\

    \midrule[.66pt]

    \parbox[t]{2mm}{\multirow{4}{*}{\rotatebox[origin=c]{90}{\texttt{flickr}}}}

    & $0$ & $0.10 \pm 0.08$ & $0.10 \pm 0.08$ & $0.11 \pm 0.09$ & $0.11 \pm 0.09$ & $0.12 \pm 0.10$ & \multirow{4}{*}{$0.11 \pm 0.09$} \\
    & $2$ & $0.10 \pm 0.08$ & $0.11 \pm 0.09$ & $0.10 \pm 0.08$ & $0.12 \pm 0.09$ & $0.11 \pm 0.09$ & \\
    & $4$ & $0.11 \pm 0.08$ & $0.11 \pm 0.09$ & $0.10 \pm 0.08$ & $0.11 \pm 0.08$ & $0.11 \pm 0.08$ & \\
    & $8$ & $0.11 \pm 0.08$ & $0.11 \pm 0.09$ & $0.10 \pm 0.08$ & $0.12 \pm 0.09$ & $0.11 \pm 0.09$ & \\

    \midrule[.66pt]

    \parbox[t]{2mm}{\multirow{4}{*}{\rotatebox[origin=c]{90}{\texttt{youtube}}}}

    & $0$ & $0.10 \pm 0.09$ & $0.10 \pm 0.09$ & $0.10 \pm 0.08$ & $0.11 \pm 0.09$ & $0.13 \pm 0.11$ & \multirow{4}{*}{$0.12 \pm 0.10$} \\
    & $2$ & $0.11 \pm 0.09$ & $0.11 \pm 0.10$ & $0.11 \pm 0.09$ & $0.16 \pm 0.12$ & $0.12 \pm 0.09$ & \\
    & $4$ & $0.10 \pm 0.08$ & $0.10 \pm 0.08$ & $0.13 \pm 0.11$ & $0.18 \pm 0.13$ & $0.13 \pm 0.09$ & \\
    & $8$ & $0.10 \pm 0.08$ & $0.10 \pm 0.08$ & $0.11 \pm 0.09$ & $0.15 \pm 0.11$ & $0.16 \pm 0.11$ & \\
    
    \bottomrule

    \end{tabular}
    \label{tab:jacc-similarity-quality75perc}
\end{table*}

\begin{table*}[h]
    \centering
    \caption{Mean and standard deviation of absolute percentage errors for Jaccard similarity estimation, with $100$ hash functions. Queries were made after inserting the first $100\%$ of the edges.}
    \begin{tabular}{lc|ccccc|c}
    \toprule
        & $k$ & $\varphi = 0.1$ & $\varphi = 0.25$ & $\varphi = 0.5$ & $\varphi = 0.75$ & $\varphi = 1$ & baseline\\
    \midrule

    \parbox[t]{2mm}{\multirow{4}{*}{\rotatebox[origin=c]{90}{\texttt{linux}}}}

    & $0$ & $0.11 \pm 0.09$ & $0.11 \pm 0.09$ & $0.09 \pm 0.08$ & $0.12 \pm 0.09$ & $0.14 \pm 0.10$ & \multirow{4}{*}{$0.08 \pm 0.07$} \\
    & $2$ & $0.09 \pm 0.07$ & $0.11 \pm 0.09$ & $0.10 \pm 0.08$ & $0.11 \pm 0.08$ & $0.15 \pm 0.11$ & \\
    & $4$ & $0.09 \pm 0.08$ & $0.11 \pm 0.08$ & $0.11 \pm 0.09$ & $0.11 \pm 0.09$ & $0.11 \pm 0.09$ & \\
    & $8$ & $0.09 \pm 0.07$ & $0.11 \pm 0.08$ & $0.10 \pm 0.08$ & $0.11 \pm 0.08$ & $0.11 \pm 0.09$ & \\

    \midrule[.66pt]
    
    \parbox[t]{2mm}{\multirow{4}{*}{\rotatebox[origin=c]{90}{\texttt{fb-wosn}}}}

    & $0$ & $0.80 \pm 0.25$ & $0.80 \pm 0.25$ & $0.80 \pm 0.24$ & $0.81 \pm 0.24$ & $0.81 \pm 0.24$ & \multirow{4}{*}{$0.80 \pm 0.24$} \\
    & $2$ & $0.80 \pm 0.24$ & $0.79 \pm 0.25$ & $0.79 \pm 0.25$ & $0.80 \pm 0.24$ & $0.81 \pm 0.24$ & \\
    & $4$ & $0.79 \pm 0.25$ & $0.79 \pm 0.25$ & $0.80 \pm 0.24$ & $0.80 \pm 0.24$ & $0.80 \pm 0.24$ & \\
    & $8$ & $0.80 \pm 0.24$ & $0.79 \pm 0.25$ & $0.79 \pm 0.25$ & $0.80 \pm 0.24$ & $0.80 \pm 0.24$ & \\

    \midrule[.66pt]
    
    \parbox[t]{2mm}{\multirow{4}{*}{\rotatebox[origin=c]{90}{\texttt{enron}}}}

    & $0$ & $0.09 \pm 0.09$ & $0.11 \pm 0.10$ & $0.12 \pm 0.10$ & $0.13 \pm 0.11$ & $0.14 \pm 0.11$ & \multirow{4}{*}{$0.09 \pm 0.09$} \\
    & $2$ & $0.09 \pm 0.09$ & $0.11 \pm 0.10$ & $0.11 \pm 0.10$ & $0.12 \pm 0.10$ & $0.13 \pm 0.11$ & \\
    & $4$ & $0.09 \pm 0.08$ & $0.10 \pm 0.09$ & $0.11 \pm 0.10$ & $0.12 \pm 0.10$ & $0.12 \pm 0.10$ & \\
    & $8$ & $0.09 \pm 0.09$ & $0.10 \pm 0.09$ & $0.10 \pm 0.09$ & $0.11 \pm 0.09$ & $0.10 \pm 0.09$ & \\

    \midrule[.66pt]

    \parbox[t]{2mm}{\multirow{4}{*}{\rotatebox[origin=c]{90}{\texttt{flickr}}}}

    & $0$ & $0.10 \pm 0.08$ & $0.10 \pm 0.08$ & $0.11 \pm 0.09$ & $0.11 \pm 0.09$ & $0.12 \pm 0.10$ & \multirow{4}{*}{$0.11 \pm 0.09$} \\
    & $2$ & $0.10 \pm 0.08$ & $0.11 \pm 0.09$ & $0.10 \pm 0.08$ & $0.12 \pm 0.09$ & $0.11 \pm 0.09$ & \\
    & $4$ & $0.11 \pm 0.08$ & $0.11 \pm 0.09$ & $0.10 \pm 0.08$ & $0.11 \pm 0.08$ & $0.11 \pm 0.08$ & \\
    & $8$ & $0.11 \pm 0.08$ & $0.11 \pm 0.09$ & $0.10 \pm 0.08$ & $0.12 \pm 0.09$ & $0.11 \pm 0.09$ & \\

    \midrule[.66pt]

    \parbox[t]{2mm}{\multirow{4}{*}{\rotatebox[origin=c]{90}{\texttt{youtube}}}}

    & $0$ & $0.10 \pm 0.09$ & $0.10 \pm 0.09$ & $0.10 \pm 0.08$ & $0.11 \pm 0.09$ & $0.13 \pm 0.11$ & \multirow{4}{*}{$0.12 \pm 0.10$} \\
    & $2$ & $0.11 \pm 0.09$ & $0.11 \pm 0.10$ & $0.11 \pm 0.09$ & $0.16 \pm 0.12$ & $0.12 \pm 0.09$ & \\
    & $4$ & $0.10 \pm 0.08$ & $0.10 \pm 0.08$ & $0.13 \pm 0.11$ & $0.18 \pm 0.13$ & $0.13 \pm 0.09$ & \\
    & $8$ & $0.10 \pm 0.08$ & $0.10 \pm 0.08$ & $0.11 \pm 0.09$ & $0.15 \pm 0.11$ & $0.16 \pm 0.11$ & \\
    
    \bottomrule

    \end{tabular}
    \label{tab:jacc-similarity-quality100perc}
\end{table*}

\end{document}